\documentclass[twoside,12pt]{article}
\usepackage{epsfig}

\def\Journal#1#2#3#4{{#1} {#2} (#4) #3 }
\def\NCA{{\em Nuovo Cimento} A}

\def\NPA{{\em Nucl. Phys.} A}

\def\NPB{{\em Nucl. Phys.} B}

\def\PLB{{\em Phys. Lett.} B}

\def\PRL{\em Phys. Rev. Lett.}
\def\PREV{\em Phys. Rev.}
\def\PREP{\em Phys. Rep.}

\def\PRD{{\em Phys. Rev.} D}
\def\PRC{{\em Phys. Rev.} C}

\def\ZPC{{\em Z. Phys.} C}

\def\RMP{{\em Rev. Mod. Phys.}}

\newcommand{\be}{\begin{equation}}
\newcommand{\ee}{\end{equation}}
\newcommand{\bea}{\begin{eqnarray}}
\newcommand{\eea}{\end{eqnarray}}

\begin{document}

\title{\bf QCD at Low Energies.}
\author{B.L.Ioffe\\
\\ Institute of Theoretical and Experimental Physics,\\
B.Cheremushkinskaya 25, 117218 Moscow,Russia}

\maketitle

\def\la{\mathrel{\mathpalette\fun <}}
\def\ga{\mathrel{\mathpalette\fun >}}
\def\fun#1#2{\lower3.6pt\vbox{\baselineskip0pt\lineskip.9pt
\ialign{$\mathsurround=0pt#1\hfil##\hfil$\crcr#2\crcr\sim\crcr}}}

\newpage

\begin{abstract}
The modern status  of basic low energy QCD parameters is reviewed.
It is demonstrated, that the recent data allows one to determine
the light quark mass ratios with an accuracy  10-15\%. The general
analysis of vacuum condensates in QCD is presented, including
those induced by external fields. The QCD coupling constant
$\alpha_s(m^2_{\tau})$ is found from the $\tau$-lepton hadronic
decay rate. The contour improved perturbation theory includes the
terms up to $\alpha^4_s$. The influence of instantons on
$\alpha_s(m^2_{\tau})$ determination is estimated. V-A spectral
functions of $\tau$-decay are used for construction of the V-A
polarization operator $\Pi_{V-A}(s)$ in the complex $s$-plane. The
operator product expansion (OPE) is used up to dimension D=10 and
the sum rules along the rays in the complex $s$-plane are
constructed. This  makes it possible to separate the contributions
of operators of different dimensions. The best values of quark
condensate and $\alpha_s \langle 0\vert \bar{q} q \vert 0
\rangle^2$ are found. The value of quark condensate is confirmed
by considering  the sum rules for baryon masses. Gluon condensate
is found in four ways: by considering of V+A polarization operator
based on the $\tau$-decay data, by studying  the sum rules for
polarization operators momenta in charmonia in the vector,
pseudoscalar and axial channels. All of these determinations are
in agreement and   result in $\langle (\alpha_s/\pi) G^2 \rangle =
0.005 \pm 0.004~ GeV^4$. Valence quark distributions in proton are
calculated in QCD using the OPE in proton current virtuality. The
quark distributions agree with those found from the deep inelastic
scattering data. The same value of gluon condensate is favoured.
\end{abstract}

\vspace{5mm}

~~\\
{\bf \large Content}\\
\\
 {\bf 1~~ Introduction}\\  {\bf 2~~ The Masses of Light Quarks}\\
 {\bf 3~~ Condensates}\\  {\bf 3.1} {\it General Properties}\\
 {\bf 3.2} {\it
Condensates, Induced  by External Fields}\\   {\bf 4~~ Test of QCD
at Low Energies on the Basis of $\tau$-decay Data}\\ {\bf 4.1}
{\it Determination of $\alpha_s(m^2_{\tau})$}\\  {\bf 4.2} {\it
Instanton Corrections}\\  {\bf 4.3} {\it Comparison with Other
Approaches}\\ {\bf 5~~ Determination of Condensates from Spectral
Functions of $\tau$-decay}\\  {\bf 5.1} {\it Determination of
Quark Condensate from V-A Spectral Function.}\\ {\bf 5.2} {\it
Determination of Condensate from V+A and V Structure Functions.}\\
{\bf 6~~ Determination of Quark Condensate from QCD Sum Rules

~~for Nucleon Mass}\\  {\bf 7~~ Gluon Condensate and Determination
of Charmed Quark Mass

~~from  Charmonium Spectrum}\\  {\bf 7.1} {\it The Method of
Moments. The Results}\\  {\bf 7.2} {\it The Attempts to Sum Up the
Coulomb-like Corrections}\\  {\bf 8~~ Valence Quark Distributions
in Nucleon at Low $Q^2$ and the Condensates.}\\ {\bf 9~~
Conclusion}\\ {\bf 10~ Acknowledgements}

\newpage

\section{Introduction}


Nowadays, it is reliably established that the true (microscopic)
theory of strong interaction is quantum chromodynamics (QCD), the
nonabelian gauge theory of interacting quarks and gluons. The main
confirmation of QCD comes from considering  the processes at high
energies and high momentum transfers, where, because of asymptotic
freedom, the high precision of theoretical calculation is achieved
and comparison with experiment confirms QCD with a very good
accuracy. In the domain of low energies and momentum transfers (by
such a domain in this paper I mean the domain of momentum
transfers $Q^2 \sim 1-5~ GeV^2$) the situation is more
complicated: the QCD coupling constant $\alpha_s$ is large,
$\alpha_s \sim 0.5-0.3$ and many loops perturbative calculations
are necessary. Unlike quantum electrodynamics (QED), the vacuum in
QCD has a nontrivial structure: due to nonperturbative effects,
non-zero fluctuations of gluonic and quark fields persist in QCD
vacuum. The nontrivial vacuum structure of QCD manifests itself in
the presence of vacuum condensates, analogous to those in condense
matter physics (for instance, spontaneous magnetization).
Therefore, $\alpha_s$ corrections and nonperturbative effects must
be correctly accounted in QCD calculation in this domain.

At lower energies and $Q^2 \la 1~ GeV^2$ analytical QCD
calculations are not reliable. The useful methods are: the chiral
effective theory, lattice calculations and various model
approaches. It is, however, very desirable to have a matching of
all these approaches with QCD calculations at $Q^2$ about $1
GeV^2$. To achieve this the knowledge of low energy QCD parameters
is necessary.

In order to fix the notations I present here the form of QCD
Lagrangian:
\be
L = i~ \sum\limits_{q} \bar{\psi}^a_q (\nabla_{\mu} \gamma_{\mu} +
i m_q)\psi^a_q - \frac{1}{4}~ G^n_{\mu \nu} G^n_{\mu \nu},
\label{1} \ee where \bea \nabla_{\mu} & = &
\partial_{\mu} - i g \frac{\lambda^n}{2} A^n_{\mu} \nonumber \\
 G^n_{\mu \nu}& = &\partial_{\mu} A^n_{\mu} - \partial_{\nu}
A^n_{\mu} + g f^{n m l} A^m_{\mu} A^l_{\nu} \label{2} \eea
$\psi^a_q$ and $A^n_{\mu}$ are quark and gluon fields, $a =
1,2,3$; $n,m,l = 1,2,...8$ are colour indeces, $\lambda^n$ and
$f^{nml}$ are Gell-Mann matrices and $f$-symbols, $m_q$ -- are
bare (current) quark masses, $q = u, d, s, c...$.

 Vacuum condensates are
very important in the elucidation of the QCD structure and in
description of hadron properties at low energies. Condensates,
particularly, quark and gluonic ones, were investigated starting
from the 70-ties.  Here, first, it should be noted the QCD sum
rule method by Shifman, Vainshtein, and Zakharov  \cite{1}, which
was based on the idea of the leading role of condensates in the
calculation of masses of the low-lying hadronic states.  In the
papers of the 70-80-ies it was assumed that the perturbative
interaction constant is comparatively small (e.g., $\alpha_s(1
GeV) \approx 0.3)$, so that it is enough to restrict oneself by
the first-order terms in $\alpha_s$ and sometimes even disregard
perturbative effects in the region of masses larger than 1 GeV. At
present it is clear that $\alpha_s$ is considerably larger
($\alpha_s(1 GeV) \sim 0.5$). In a number of cases there appeared
the results of perturbative calculations in order $\alpha^2_s$ and
$\alpha^3_s$. New, more precise experimental data at low energies
had been obtained.

This review presents the modern status of QCD at low energies.  In
Chapter 2 the values of light quark masses are discussed. Chapter
3 contains the definition of condensates and the description of
their general properties. In Chapter 4 the QCD coupling constant
$\alpha_s$ is determined from the data on hadronic $\tau$-decay
and its evolution with $Q^2$ (in 4-loops approximation) is given.
In Chapter 5 quark and gluon condensates are found from the
$\tau$-decay data on V-A, V+A and V correlators. The sum rules for
nucleon mass with account of  $\alpha_s$ corrections  are analyzed
in Chapter 6 and it is shown, that they are well satisfyed  at the
same value of quark condensate, which was found from the
$\tau$-decay data.  Various ways of gluon condensate
determination: a) from $V+A$ correlators; b) from charmonium sum
rules are considered in Chapter 7. The QCD sum rules for valence
quark distributions in nucleon are presented in Chapter 8, valence
$u$- and $d$-quark distributions at low $Q^2$ were found and the
restriction on condensates were obtained. Finally, chapter 9
summarizes the state of the art of low energy QCD.

\bigskip

\section{The masses of light quarks}

The $u, d, s$ quark masses had been first estimated by Gasser and
Leutwyler about 30 years ago: it was demonstrated that $m_u, m_d
\sim 5~ MeV$ and $m_s \sim 100 MeV$ \cite{2,3}. In 1977 Weinberg
\cite{4}, using partial conservation of axial current and Dashen
theorem \cite{5} to account for electromagnetic selfenergies of
mesons had proved, that the ratios $m_u/m_d$ and $m_s/m_d$ may be
expressed through $K$ and $\pi$ masses:
\be
\frac{m_u}{m_d} = \frac{m^2_{K^+} - m^2_{K^0} + 2 m^2_{\pi^0} -
m^2_{\pi^+}}{m^2_{K^0} - m^2_{K^+} + m^2_{\pi^+}} \label{3} \ee
\be
\frac{m_s}{m_d} = \frac{m^2_{K^0} + m^2_{K^+} +
m^2_{\pi^+}}{m^2_{K^0} - m^2_{K^+} + m^2_{\pi^+}} \label{4} \ee
Numerically, (\ref{3}) and (\ref{4}) are equal
\be
\frac{m_u}{m_d} = 0.56 ~~~~~~~ \frac{m_s}{m_d} = 20.1 \label{5}
\ee Basing on consideration of mass splitting in baryon octet
Weinberg assumed, that $m_s = 150 MeV$ at the scale of about 1
GeV. Then
\be
m_u = 4.2 MeV, ~~~ m_d = 7.5 MeV, ~~~ m_s = 150 MeV \label{6} \ee
at 1 GeV. The large $m_s/m_d$ ratio explains the large mass
splitting in pseudoscalar meson octet. For $m^2_{K^+}/m^2_{\pi^+}$
we have $[\bar{m} = (m_u + m_d)/2]$
\be
\frac{m^2_{K^+}}{m^2_{\pi^+}} = \frac{m_s + \bar{m}}{2 \bar{m}} =
13 \label{7} \ee in a perfect agreement with experiment. The ratio
$m^2_{\eta}/m^2_{\pi}$ expressed in terms of quark mass ratios is
also in a good agreement with experiment.

The ratios (\ref{3}),(\ref{4}) were obtained in the first order in
quark masses. Therefore,  their accuracy is of order of accuracy
of SU(3) symmetry, i.e. about 20\%.

In \cite{6} it was demonstrated that there is a relation valid in
the second order in quark masses
\be
\Biggl (\frac{m_u}{m_d} \Biggr )^2 + \frac{1}{Q^2} \Biggl
(\frac{m_s}{m_d} \Biggr )^2 = 1 \label{8} \ee Using Dashen theorem
for electromagnetic selfenergies of $\pi$ and $K$-meson, one may
express $Q$ as
\be
Q^2_D = \frac{(m^2_{K^0} + m^2_{K^+} - m^2_{\pi^+} +
m^2_{\pi^0})(m^2_{K^0} + m^2_{K^+} - m^2_{\pi^+} - m^2_{\pi^0})}{4
m^2_{\pi^0}(m^2_{K^0} - m^2_{K^+} + m^2_{\pi^+} - m^2_{\pi^0})}
\label{9} \ee Numerically, $Q_D$ is equal: $Q_D = 24.2$. However,
Dashen theorem is valid in the first order in quark masses. The
electromagnetic mass difference of $K$-mesons calculated in
\cite{Perez} by using Cottingham formula and in \cite{Bijnens} by
large $N_c$ approach increased $\Delta m_k = (M_{K^+} -
M_{K^0})_{e.m.}$ from Dashen value $\Delta m_k = 1.27$~MeV to
$\Delta m_k = 2.6$~ MeV and, correspondingly, decreased $Q$ to $Q
= 22.0 \pm 0.6$. The other way to find $Q$ is from $\eta\to
\pi^+\pi^-\pi^0$ decay, using the chiral effective theory.
Unfortunately, the next to leading corrections are large in this
approach \cite{J.Gasser}, what makes uncertain the accuracy of the
results. It was found  from the $\eta \to \pi^+ \pi^- \pi^0$ decay
data with the account of interaction in the final state: $Q = 22.4
\pm 0.9$ \cite{Kambor}, $Q = 22.7 \pm 0.8$ \cite{Anisovich} and $Q
= 22.8 \pm 0.4$ \cite{Sopov} (the latter from the Dalitz plot).
So, the final conclusion is that $Q$ is in the interval $21.5 < Q
< 23.5$.
 (It must be mentioned that the experiment,
  where $\Gamma(\eta \to
2 \gamma)$ and, consequently, $\Gamma(\eta \to \pi^+ \pi^- \pi^0)$
were measured by the Primakoff effect, is absent in the last
edition of the Particle data \cite{10}, while it persisted in the
previous ones. See \cite{7} for the review.)  The ratio $\gamma =
m_u/m_d$ can also be found from the ratio of $\psi^{\prime} \to
(J/\psi)\eta$ and $\psi^{\prime} \to (J/\psi)\pi^0$ decays
\cite{8,9} . In \cite{8} it was proved that
\be
r = \frac{\Gamma(\psi^{\prime} \to J/\psi +
\pi^0)}{\Gamma(\psi^{\prime} \to J/\psi + \eta)} = 3 \Biggl
(\frac{1 - \gamma}{1 + \gamma}\Biggr )^2 \Biggl
(\frac{m_{\pi}}{m_{\eta}}\Biggr )^4~\Biggl
(\frac{p_{\pi}}{p_{\eta}} \Biggr )^3 \label{10} \ee where
$p_{\pi}$ and $p_{\eta}$ are the pion and $\eta$ momenta in
$\psi^{\prime}$ rest frame. Eq.(\ref{10}) is valid in the first
order in quark mass. The Particle Data Group \cite{10} gives
\be
r_{exp} = (3.04 \pm 0.71)\cdot 10^{-2} \label{11} \ee In the
recent CLEO Collaboration experiment \cite{Collab} it was found:
$r_{exp} = (4.01 \pm 0.45)\cdot 10^{-2}$. Averaging these two
experimental numbers,
  assuming
the theoretical uncertainty in (\ref{10}) as 30\% and adding in
quadratures the theoretical and experimental errors, we get from
(\ref{10})
\be
\gamma = \frac{m_u}{m_d} = 0.407 \pm 0.060 \label{12} \ee The
value close to (\ref{12}) was found recently in
\cite{Nasrallah}.The substitution of (\ref{12}) into (\ref{8})
with the account of the mentioned above uncertainty of $Q$,
results in
\be
\frac{m_s}{m_d} = 20.8 \pm 1.3 \label{13} \ee The value (\ref{12})
is slightly lower, then the lowest order result (\ref{5}),
(\ref{13}) agrees with it. The values (\ref{12}),(\ref{13})  are
in agreement with recent lattice calculations \cite{11}.

The calculation of absolute values of quark masses is a more
subtle problem. First of all, the masses are scale dependent. In
perturbation theory their scale dependence is given by the
renormalization group equation: \be\frac{dm(\mu)}{m(\mu)}=-\gamma
[~\alpha_s(\mu)~]\frac{d\mu^2}{\mu^2}=-\sum^{\infty}_{r=1}
\gamma_ra^r(\mu^2)\frac{d\mu^2}{\mu^2}\label{14}\ee In (\ref{14})
$a=\alpha_s/\pi, \gamma_1=1, \gamma_2=91/24, \gamma_3=10.48$ for 3
flavours in  $\overline{MS}$ scheme \cite{12}. In the first order
in $\alpha_s$ it follows from (\ref{14}) that: \be
\frac{m(Q^2)}{m(\mu^2)}=\Biggl
[\frac{\alpha_s(\mu^2)}{\alpha_s(Q^2)}\Biggr
]^{\gamma_m},\label{15}\ee where $\gamma_m=-4/9$ is the quark mass
anomalous dimension. There is no good  convergence of the series
(\ref{14}) below $\mu=2~ GeV (\alpha_s(2~ GeV)=0.31)$. The recent
calculations of $m_s$ by QCD sum rules \cite{13}, from the
$\tau$-decay data \cite{14,15} and on lattice \cite{11,16}, are in
a not quite good agreement with one another. The mean value
estimated in \cite{10} is: $m_s(2~ GeV)\approx 105~ MeV$ with an
accuracy of about 20\%. By taking $m_s(1~GeV)/m_s(2~ GeV)=1.35$ we
have then: $m_s(1~GeV)\approx 142~ MeV$ and, according to
(\ref{12}),(\ref{13}), $m_d(1~ GeV)= 6.8~ MeV, m_u(1~ GeV)=2.8~
MeV$. The difference $m_d-m_u$ is equal to:  $m_d-m_u=4.0\pm 1.0~
MeV$. This value agrees with one found by QCD sum rules from
baryon octet mass splitting \cite{17} and $D$ and $D^*$ isospin
mass differences \cite{18}, $m_d-m_u=3\pm 1~ MeV$. For $m_u + m_d$
we have $m_u + m_d = 9.6 \pm 2.5~ MeV$ in comparison with $m_u =
m_d = 12.8 \pm 2.5~ MeV$ found in \cite{Prades}.
 For
completeness I present here also the value of $m_c(m_c)$ (see
below, Sec.7.1)
\be
m_c(m_c)=1.275 \pm 0.015~GeV\label{16}\ee

\bigskip

\section{Condensates}

\subsection{\it General properties}


In QCD (or in a more general case, in quantum field theory) by
condensates one mean the vacuum mean values $\langle 0 \vert O_i
\vert 0 \rangle$  of the local (i.e.  taken at a single point of
space-time) of the operators $O_i(x)$, which arise due to
nonperturbative effects. The latter point is very important and
needs clarification. When determining vacuum condensates one
implies the averaging only over nonperturbative fluctuations.  If
for some operator $O_i$ the non-zero vacuum mean value appears
also in the perturbation theory, it should not be taken into
account in determination of the condensate -- in other words, when
determining condensates the perturbative vacuum mean values should
be subtracted in calculation of the vacuum averages. One more
specification is necessary. The perturbation theory series in QCD
are asymptotic series.  So, vacuum mean operator values may appear
due to one or another summing of asymptotic series.  The vacuum
mean values of such kind are commonly to be referred to vacuum
condensates.

In quantum field theory it is assumed, that vacuum correlators
$\Pi_{AB}(x,y)$ in coordinate space of any two local operators
$A(x)$, $B(y)$

$$ \Pi_{AB}(x,y) = \langle 0\vert T\left \{A(x), B(y) \right \}
\vert 0 \rangle $$ at space-like $(x-y)^2 \leq 0$, and small
$x-y~~(x - y \to 0)$ may be represented as an operator product
expansion (OPE) series

$$ \Pi_{AB} (x-y) = \sum\limits_{i}~a_i (x-y) \langle 0 \vert O_i
(0) \vert 0 \rangle, $$ where $a_i(x-y)$ are called the
coefficient functions and are given by perturbation theory. (The
strict proofs of this statement were obtained only in perturbation
theory and for some models). Here, again one must take care of
separation of perturbative and nonperturbative parts in the
definition of condensates. The perturbation expansion for
$a_i(x-y)$ is an asymptotic series and the terms which arise by
summing of such series may be interpreted as contributions of
higher dimension operators. $a_i(x-y)$ may be infra-red divergent.
This is a signal of appearance of an additional condensate in OPE.
Also, probably, OPE for $\Pi_{AB}(x-y)$ is asymptotic series. In
order to avoid all these problems in practical calculations, it is
necessary to require a good convergence of OPE and perturbation
series in the domain of interest.

 Separation of perturbative and nonperturbative contribution into
vacuum mean values has some arbitrariness. Usually \cite{19,20},
this arbitrariness is avoided by introducing some normalization
point $\mu^2$ ~ ($\mu^2 \sim 1 GeV^2$). Integration over  momenta
of virtual quarks and gluons in the region below $\mu^2$ is
referred to condensates, above $\mu^2$ -- to perturbative theory.
In such a formulation condensates depend on the normalization
point $\mu$: ~ $\langle 0 \vert O_i \vert 0 \rangle = \langle 0
\vert O_i \vert 0 \rangle _{\mu}$. Other methods for determination
of condensates are also possible (see below Sec.5.2).

In perturbation theory, there appear corrections to  condensates
as a series in the coupling constant $\alpha_s(\mu)$:
\be
\langle 0 \vert O_i \vert 0 \rangle_Q = \langle 0 \vert O_i \vert
0 \rangle_{\mu} \sum\limits^{\infty}_{n=0}~ C^{(i)}_n (Q, \mu)
\alpha^n_s (\mu) \label{17}\ee The running coupling constant
$\alpha_s$ at the right-hand part of (\ref{17}) is normalized at
the point $\mu$. The left-hand part of (\ref{17}) represents the
value of the condensate normalized at the point $Q$. Coefficients
$C^{(i)}_n(Q, \mu)$ may have logarithms $ln Q^2/\mu^2$ in powers
up to $n$ for $C^{(i)}_n$. Summing up of the terms with highest
powers of logarithms leads to appearance of the so-called
anomalous dimension of operators, so that in general form it can
be written
\be
\langle 0 \vert O_i \vert 0 \rangle_Q = \langle 0 \vert O_i \vert
0 \rangle_{\mu} \Biggl (\frac{\alpha_s(\mu)}{\alpha_s(Q)} \Biggr
)^{\gamma_i} \sum\limits^{\infty}_{n=0}~ c^{(i)}_n (Q, \mu)
\alpha^n_s (\mu), \label{18}\ee where $\gamma_i$ - are anomalous
dimensions (numbers), and $c^{(i)}_n$ have already no leading
logarithms. If there exist several operators of the given
(canonical) dimension, then their mixing is possible in
perturbation theory. Then the relations (\ref{17}),(\ref{18})
become matrix.

In their physical properties condensates in QCD have much in
common with condensates appearing in condensed matter physics:
such as superfluid liquid (Bose-condensate) in liquid $^4He$,
Cooper pair condensate in superconductor, spontaneous
magnetization in magnetic etc. That is why, analogously to effects
in the physics of condensed  matter, it can be expected that if
one considers QCD at finite temperature $T$, with $T$ increasing
at some $T = T_c$ there will be phase transition and condensates
(or a part of them) will be destroyed. Particularly, such a
phenomenon must hold for condensates responsible for spontaneous
symmetry breaking -- at $T = T_c$ they should vanish and symmetry
must be restored. (In principle, surely, QCD may have a few phase
transitions).

Condensates in QCD are divided into two types: conserving and
violating chirality. As was demonstrated in previous Chapter, the
masses of light quarks $u, d, s$ in the QCD Lagrangian are small
comparing with the characteristic scale of hadronic masses $ M
\sim 1~ GeV$. In neglecting  light quark masses the QCD Lagrangian
becomes chiral-invariant: left-hand and right-hand (in chirality)
light quarks do not interact with each other, both vector and
axial currents are conserved (except for flavour-singlet axial
current, non-conservation of which is due to anomaly). The
accuracy of light quark masses neglect corresponds to the accuracy
of isotopical symmetry, i.e. a few per cent in the case of $u$ and
$d$ quarks and of the accuracy of SU(3) symmetry, i.e. 10-15 \% in
the case of $s$-quarks. In the case of condensates violating
chiral symmetry, perturbative vacuum mean values are proportional
to light quark masses and are zero within $m_u = m_d = m_s = 0$.
So, such condensates are determined in the theory much better than
those conserving chirality and, in principle, may be found
experimentally with a higher accuracy.

Among chiral symmetry violating condensates of the most importance
is the quark condensate $\langle 0 \vert \bar{q} q \vert 0
\rangle$~ ($q = u, d$ are the fields of $u$ and $d$ quarks).
$\langle 0 \vert \bar{q} q \vert 0 \rangle$ may be written in the
form
\be
\langle 0 \vert \bar{q} q \vert 0 \rangle = \langle 0 \vert
\bar{q}_L q _R + \bar{q}_R q_L \vert 0 \rangle\label{19} \ee where
$q_L, q_R$ are the fields of left-hand and right-hand (in
chirality) quarks. As follows from (\ref{19}), the non-zero value
of quark condensate means the transition of left-hand quark fields
into right-hand ones and its not a small value would mean the
chiral symmetry violation in QCD. (If chiral symmetry is not
violated spontaneously, then at small $m_u, m_d$ ~~ $\langle 0
\vert \bar{q} q \vert 0 \rangle \sim m_u, m_d$). By virtue of
isotopical invariance
\be
\langle 0 \vert \bar{u} u \vert 0 \rangle = \langle 0 \vert
\bar{d} d \vert 0 \rangle\label{20} \ee For quark condensate there
holds the Gell-Mann-Oakes-Renner relation \cite{21}
\be
\langle 0 \vert \bar{q} q \vert 0 \rangle = - \frac{1}{2}~
\frac{m^2_{\pi} f^2 _{\pi}}{m_u + m_d}\label{21} \ee Here
$m_{\pi}, f_{\pi}$ are the mass and constant of $\pi^+$-meson
decay ($m_{\pi} = 140 MeV, ~ f_{\pi} = 131 MeV$), ~ $m_u$ and
$m_d$ are the masses of $u$ and $d$-quarks. Relation (\ref{21}) is
obtained in the first order in $m_u, m_d, m_s$ (for its derivation
see, e.g. \cite{22}). To estimate the value of quark condensate
one may use the values of quark masses $m_u+m_d=9.6~ MeV$,
presented in Sec.2. Substituting these values into (\ref{21}) we
get
\be
\langle 0 \vert \bar{q} q \vert 0 \rangle = - (260~
MeV)^3\label{22} \ee The value (\ref{6}) has characteristic
hadronic scale. This shows that chiral symmetry which is fulfilled
with a good accuracy in the light quark lagrangian ($m_u, m_d/M
\sim 0.01$), is spontaneously violated on hadronic state spectrum.

 An other argument in the favour of spontaneous violation of
chiral symmetry in QCD is the existence of massive baryons.
Indeed, in the chiral-symmetrical theory all fermionic states
should be either massless or parity-degenerated. Obviously,
baryons, in particular, nucleon do not possess this property. It
can be shown \cite{23,22},
 that both these phenomena -- the presence of the
chiral symmetry violating quark condensate and the existence of
massive baryons are closely connected with each other. According
to the Goldstone theorem, the spontaneous symmetry violation leads
to appearance of massless particles in the physical state spectrum
-- of Goldstone bosons. In QCD Goldstone bosons can be identified
 with a $\pi$-meson triplet within $m_u, m_d \to 0$, ~ $m_s \not=
0$ (SU(2)-symmetry) or with an octet of pseudoscalar mesons
($\pi$, $K, \eta$) within the limit $m_u, m_d, m_s \to 0$
(SU(3)-symmetry). The presence of Goldstone bosons in QCD makes it
possible to formulate the low-energy chiral effective theory of
strong interactions (see reviews \cite{24},\cite{25},\cite{22}).

Quark condensate may be considered as an order parameter in QCD
corresponding to spontaneous violation of the chiral symmetry. At
the temperature of restoration of the chiral symmetry $T = T_c$ it
must vanish. The investigation of the temperature dependence of
quark condensate in the chiral effective theory \cite{26} shows
that  $\langle 0 \vert \bar{q} q \vert 0 \rangle$ vanishes at $T =
T_c \approx 150-200 MeV$. Similar indications were obtained also
in the lattice calculations \cite{27}.

Thus, the quark condensate: 1) has the lowest dimensions (d=3) as
compared with other condensates in QCD; 2) determines masses of
usual (nonstrange) baryons; 3) is the order parameter in the phase
transition between the phases of violated and restored chiral
symmetry. These three facts determine its important role in the
low-energy hadronic physics.

Let us estimate the accuracy of numerical value of (\ref{22}). The
quark condensate, as well as quark masses depend on the
normalization point and have anomalous dimensions equalling to
$\gamma_{\bar{q}q} = -\gamma_m = \frac{4}{9}$. In (\ref{22}) the
normalization point $\mu$ was taken $\mu \simeq 1~ GeV$.  The
Gell-Mann-Oakes-Renner relation is derived up to correction terms
linear in quark masses. In the chiral effective theory it is
possible to estimate the correction terms and, thereby, the
accuracy of equation (\ref{21}) appears to be of order 10\%.  The
accuracy of the above taken value $m_u + m_d = 9.6~ MeV$ which
enters (\ref{21}) seems to be of order 20\%. The value of the
quark condensate may be also found from the sum rules for proton
mass (see Chapt.6) as well as from structure functions at
$\tau$-decay (Chapt.5).  The quark condensate of strange quarks is
somewhat different from $\langle 0 \vert \bar{u}u \vert 0
\rangle$. In \cite{23} it was obtained
\be
\langle 0 \vert \bar{s} s \vert 0 \rangle/ \langle 0 \vert \bar{u}
u \vert 0 \rangle = 0.8 \pm 0.1 \label{23}\ee The next in
dimension (d = 5) condensate which violates chiral symmetry is
quark gluonic one:
\be
g \langle 0 \vert \bar{q} \sigma_{\mu \nu} \frac{\lambda^n}{2}
G^n_{\mu \nu} q \vert 0 \rangle \equiv m^2_0 \langle 0 \vert
\bar{q} q \vert 0\rangle \label{24}\ee Here $G^n_{\mu \nu}$ - is
the gluonic field strength tensor, $\lambda^n$ - are the Gell-Mann
matrices, $\sigma_{\mu \nu} = (i/2)(\gamma_{\mu} \gamma_{\nu} -
\gamma_{\nu}\gamma_{\mu}$). The value of the parameter $m^2_0$ was
found in \cite{28} from the sum rules for baryonic resonances
\be
m^2_0 = 0.8\pm 0.2~ GeV^2\label{25} \ee The same value of $m^2_0$
was found from the analysis of $B$-mesons by QCD sum rules
\cite{29}, close to (\ref{25}) value of $m^2_0=1.0~GeV^2$ was
calculated in the model of field correlators \cite{Giacomo}. The
anomalous dimension of the operator in (\ref{24}) is small
\cite{30}. Therefore the anomalous dimension of $m^2_0$ is
approximately equal  to $\gamma_m=-4/9$.

 Consider now condensates conserving
chirality. Of fundamental role here is the gluonic condensate of
the lowest dimension:
\be
\langle 0 \vert \frac{\alpha_s}{\pi} G^n_{\mu \nu} G^n_{\mu \nu}
\vert 0\rangle \label{26}\ee Since gluonic condensate is
proportional to the vacuum mean value of the trace of the
energy-momentum tensor $\theta_{\mu \nu}$ its anomalous dimension
is zero. The existence of gluonic condensate had been first
indicated by Shifman, Vainshtein, and Zakharov \cite{1}. They had
also obtained its numerical value from the sum rules for
charmonium:
\be
\langle 0 \vert \frac{\alpha_s}{\pi} G^n_{\mu \nu} G^n_{\mu \nu}
\vert 0\rangle =0. 012 GeV^4 \label{27}\ee As was shown by the
same authors, the nonzero and positive value of gluonic condensate
means, that the vacuum energy is negative in QCD: vacuum energy
density in QCD is given by $\varepsilon = -(9/32) \langle 0 \vert
(\alpha_s/\pi) G^2 \vert 0 \rangle$. The persistence of quark
field in vacuum destroys (or suppresses) the condensate.
 Therefore, if quark is
embedded into vacuum, this results in its excitation, i.e, in
increasing of energy. Thereby, it become possible to explain the
bag model  in QCD: in the domain around quark there appears an
excess of energy, which is treated as the energy density $B$ in
the bag model. (Although, the magnitude of $B$, does not,probably,
agree with the value of $\varepsilon$ which follows from
(\ref{27})). In ref.\cite{1} perturbative effects were taken into
account only in the order $\alpha_s$, the value for $\alpha_s$
being taken about two times smaller as the modern one. Later many
attempts were made to determine the value of gluonic condensate by
studying various processes and by applying various methods.  But
the results of different approaches were inconsistent with each
other and with (\ref{27}) and sometimes the difference was even
very large -- the values of condensate appeared to be by a few
times larger. All of this requires to reanalyse  the methods  of
$\langle 0 \vert \frac{\alpha_s}{\pi} G^2 \vert 0 \rangle$
determination basing on modern values of $\alpha_s$ that will be
done in Sections 7,8.

The d=6 gluonic condensate is of the form
\be
g^3 f^{abc} \langle 0 \vert G^a_{\mu \nu} G^b_{\nu \lambda}
G^c_{\lambda \mu} \vert 0 \rangle, \label{28}\ee $(f^{abc}$ - are
structure constants of SU(3) group). There are no reliable methods
to determine it from experimental data. There is only an estimate
\cite{31} which follows from the model of deluted instanton gas:
\be
g^3 f^{abc} \langle 0 \vert G^a_{\mu \nu} G^b_{\nu \lambda}
G^c_{\lambda \mu} \vert 0 \rangle = \frac{4}{5} (12 \pi^2)
\frac{1}{\rho^2_c} \langle 0 \vert \frac{\alpha_s}{\pi} G^2_{\mu
\nu} \vert 0 \rangle, \label{29}\ee where $\rho_c$ is the
instanton effective radius in the given model (for estimation one
may take $\rho_c \sim (1/3 - 1/2) fm)$.

The general form of d=6 condensates built from quark fields is:
\be
\alpha_s \langle 0 \vert \bar{q}_i O_{\alpha} q_i \cdot \bar{q}_k
O_{\alpha} q_k \vert 0 \rangle \label{30}\ee where $q_i, q_k$ are
quark fields of $u, d, s$ quarks, $O_{\alpha}$ - are Dirac and
$SU(3)$ matrices. Following \cite{1}, Eq.(\ref{30}) is usually
factorized: in the sum over intermediate state  in all channels
(i.e,  if necessary, after Fierz-transformation) only vacuum state
is taken into account. The accuracy of such approximation $\sim
1/N^2_c$, where $N_c$ is the number of colours i.e.$\sim 10\%$.
After factorization Eq.(\ref{30}) reduces to
\be
\alpha_s \langle 0 \vert \bar{q} q \vert 0 \rangle^2,\label{31}
\ee if $q=u,d.$ The anomalous dimension of (\ref{31}) is -- 1/9
and it can be approximately put to be zero. And finally, d=8 quark
condensates assuming factorization reduce to
\be
\alpha_s \langle 0 \vert \bar{q} q \vert 0 \rangle  \cdot m^2_0
\langle 0 \vert  \bar{q} q \vert 0 \rangle \label{32}\ee (The
notation of (\ref{24}) is used). It should be noted, however, that
the factorization procedure in the d=8 condensate case is not
quite certain. For this reason, it is necessary to require their
contribution to be small.

There are few gluon and quark-gluon condensates of dimension 8.
(The full list of them is given in \cite{32}.) As a rule,
factorization hypothesis is used for their calculation. The other
way to estimate the values of these condensate is to use the
dilute instanton gas model. However, the latter for some
condensates gives the results (at accepted values of instanton gas
model parameters) by one  order of magnitude larger, than the
factorization  method. The arguments were presented \cite{33},
that instanton gas model overestimates the values of $d=8$ gluon
condensate. Therefore, the estimates based on factorization
hypothesis are more reliable here.

The violation of factorization hypothesis is more strong for
higher dimension condensates. So, this hypothesis may be used only
for their estimations by the order of magnitude.

\subsection{\it Condensates, induced by external fields}

 The meaning of such condensates
can be easily understood by comparing with analogous phenomena in
the physics of condensed matter. If the above considered
condensates can be compared, for instance with ferromagnetics,
where magnetization is present even in the absence of external
magnetic field, condensates induced by external field are similar
to dia- or paramagnetics. Consider the case of the constant
external electromagnetic field $F_{\mu \nu}$. In its presence
there appears a condensate induced  by external field (in the
linear approximation in $F_{\mu \nu}$):
\be
\langle 0 \vert \bar{q} \sigma_{\mu \nu} q \vert 0 \rangle_F = e_q
\chi F_{\mu \nu} \langle 0 \vert \bar{q} q \vert 0 \rangle
\label{33}\ee As was shown in ref.\cite{34}, in a good
approximation $\langle 0 \vert \bar{q} \sigma_{\mu \nu} q \vert 0
\rangle_F$ is proportional to $e_q$ - the charge of quark $q$.
Induced by the field vacuum expectation value $\langle 0 \vert
\bar{q} \sigma_{\mu \nu} q \vert 0 \rangle_F$ violates chiral
symmetry. So, it is natural to separate $\langle 0 \vert \bar{q}q
\vert 0 \rangle$ as a factor in eq.(\ref{33}).  The universal
quark flavour independent quantity $\chi$ is called magnetic
susceptibility of quark condensate. Its numerical value had been
found in \cite{35} using a special sum rule:
\be
\chi = -(5.7 \pm 0.6) GeV^2\label{34} \ee Another example is
external constant axial isovector field $A_{\mu}$, the interaction
of which with light quarks is described by the Lagrangian
\be
L^{\prime} = (\bar{u} \gamma_{\mu} \gamma_5 u - \bar{d}
\gamma_{\mu} \gamma_5 d) A_{\mu}\label{35} \ee In the presence of
this field there appear induced by it condensates:
\be
\langle 0 \vert \bar{u} \gamma_{\mu} \gamma_5 u \vert 0 \rangle_A
= - \langle 0 \vert \bar{d} \gamma_{\mu} \gamma_5 d\vert 0 \rangle
_A = f^2_{\pi} A_{\mu} \label{36}\ee where $f_{\pi} = 131 MeV$ is
the constant of $\pi \to \mu \nu$ decay. The right-hand part of
eq.(\ref{36}) is obtained assuming $m_u, m_d \to 0,$~ $m^2_{\pi}
\to 0$ and follows directly from consideration of the polarization
operator of axial currents $\Pi^A_{\mu \nu}(q)$  in the limit $q
\to 0$, when because of axial current conservation the nonzero
contribution into $\Pi^A_{\mu \nu}(q)_{q \to 0}$ emerges only from
one-pion intermediate state. Eq.(\ref{36}) was used to calculate
the axial coupling constant in $\beta$-decay $g_A$ \cite{36}. An
analogous to (\ref{36}) relation holds in the case of octet axial
field. Of special interest is the condensate induced by singlet
(in flavours) constant axial field
\be
\langle 0 \vert j^{(0)}_{\mu 5} \vert 0 \rangle = 3 f^2_0
A^{(0)}_{\mu} \label{37}\ee
\be
j^{(0)}_{\mu 5} = \bar{u} \gamma_{\mu} \gamma_5 u + \bar{d}
\gamma_{\mu} \gamma_5 d +  \bar{s} \gamma_{\mu} \gamma_5 s
\label{38}\ee and the Lagrangian of interaction with external
field has the form
\be
L^{\prime} = j^{(0)}_{\mu 5} A^{(0)}_{\mu} \label{39}\ee Constant
$f_0$ cannot be calculated by the method used when deriving
eq.(\ref{36}), since singlet axial current is not conserved
because of anomaly and the singlet pseudoscalar meson
$\eta^{\prime}$ is not Goldstone one. The constant $f^2_0$ is
proportional to topological susceptibility of vacuum \cite{37}
\be
f^2_0 = \frac{4}{3} N^2_f \chi^{\prime} (0), \label{40}\ee where
$N_f$ is the number of light quarks, $N_f = 3$, and the
topological susceptibility of the vacuum $\chi(q^2)$ is defined as
\be
\chi(q^2) = i \int~ d^4 xe^{iqx} \langle 0 \vert T {Q_5(x),~
Q_5(0)} \vert 0 \rangle \label{41}\ee
\be
Q_5(x) = \frac{\alpha_s}{8 \pi} G^n_{\mu \nu} (x) \tilde{G}^n_{\mu
\nu} (x), \label{42}\ee where $\tilde{G}^n_{\mu\nu}$ is dual to
$G^n_{\mu\nu}:$
$\tilde{G}^n_{\mu\nu}=(1/2)\varepsilon_{\mu\nu\lambda\sigma}
G^n_{\lambda\sigma}.$ Using the QCD sum rule, one may relate
$f^2_0$ with the part of proton spin $\Sigma$, carried by quarks
in polarized $ep$ (or $\mu p$) scattering \cite{37}. The value of
$f^2_0$ was found from the selfconsistency condition of the
obtained sum rule (or from the experimental value of $\Sigma$):
\be
f^2_0 = (2.8 \pm 0.7) \cdot 10^{-2} GeV^2\label{43} \ee The
related to it value of the derivative at $q^2 = 0$ of vacuum
topological susceptibility $\chi^{\prime}(0)$, (more precisely,
its nonperturbative part) is equal to:
\be
\chi^{\prime} (0) = (2.3 \pm 0.6) \cdot 10^{-3} GeV^2
\label{44}\ee The value $\chi^{\prime} (0)$ is of essential
interest for studying properties of vacuum in QCD.

\bigskip

\section{Test of QCD at low energies on the basis of
$\tau$-decay data}

\subsection{\it Determination of $\alpha_s(m^2_{\tau})$}


Collaborations ALEPH \cite{38}, OPAL \cite{39} and CLEO \cite{40}
had measured with a good accuracy the relative probability of
hadronic decays of $\tau$-lepton $R_{\tau} = B(\tau \to \nu_{\tau}
+ hadrons)/B(\tau \to \nu_{\tau} e\overline{\nu}_e)$, the vector
$V$ and axial $A$ spectral functions. Below I present the results
of the theoretical analysis of these data basing on the operator
product expansion (OPE) in QCD \cite{41,42} (see also
\cite{43,44}). In the perturbation theory series the terms up to
$\alpha^4_s$ will be taken into account, in OPE -- the operators
up to dimension 8. I restrict myself to the case of equal to zero
total hadronic strangeness.

Consider the polarization operator of hadronic currents
\be
\Pi^J_{\mu \nu} = i~ \int~ e^{iqx} \langle T J_{\mu} (x) J_{\nu}
(0)^{\dag} \rangle dx = (q_{\mu} q_{\nu} - \delta_{\mu \nu} q^2)
\Pi^{(1)}_J (q^2) + q_{\mu} q_{\nu} \Pi^{(0)}_J (q^2),\label{45}
\ee $$ \mbox{where} ~~~~ J = V,A; ~~~ V_{\mu} = \bar{u}
\gamma_{\mu} d, ~~~ A_{\mu} = \bar{u} \gamma_{\mu} \gamma_5 d. $$
The spectral functions measured in $\tau$-decay are imaginary
parts of $\Pi^{(1)}_J(s)$ and $\Pi^{(0)}_J(s)$, ~ $s = q^2$
\be
v_1/a_1(s) = 2\pi Im \Pi^{(1)}_{V/A} (s + i 0), ~~~ a_0(s) = 2 \pi
Im \Pi^{(0)}_A (s + i0)\label{46} \ee Functions $\Pi^{(1)}_V(q^2)$
and $\Pi^{(0)}_A(q^2)$ are analytical functions in the $q^2$
complex plane  with a cut along the right-hand semiaxis starting
from $4 m^2_{\pi}$ for $\Pi^{(1)}_V(q^2)$ and $9m^2_{\pi}$ for
$\Pi^{(0)}_A( q^2)$. Function $\Pi^{(1)}_A(q^2)$ has kinematical
pole at $q^2 = 0$, since the physical combination, which have no
singularities is $\delta_{\mu \nu} q^2 \Pi^{(1)}_A (q^2)$. Because
of axial current conservation  in the limit of massless quarks
this kinematical pole is related
 to one-pion state contribution into
$\Pi_A(q)$, which has the form \cite{41}
\be
\Pi^A_{\mu \nu}(q)_{\pi} = -\frac{f^2_{\pi}}{q^2} (q_{\mu} q_{\nu}
- \delta_{\mu \nu} q^2) - \frac{m^2_{\pi}}{q^2} q_{\mu} q_{\nu}
\frac{f^2_{\pi}}{q^2 - m^2_{\pi}}\label{47} \ee The chiral
symmetry violation may result in corrections of order
$f^2_{\pi}(m^2_{\pi}/m^2_{\rho})$ in $\Pi^{(1)}_A(q^2)$
($m_{\rho}$ is the characteristic hadronic mass), i.e. in the
theoretical uncertainty in the magnitude of the residue of
kinematical pole in $\Pi^{(1)}_A(q^2)$ of order $\Delta
f^2_{\pi}/f^2_{\pi} \sim m^2_{\pi}/m^2_{\rho}$.

 Consider first the
ratio of the total probability of hadronic decays of
$\tau$-leptons into states with zero strangeness to the
probability of $\tau \to \nu_{\tau} e \overline{\nu}_e$. This
ratio is given by the equality \cite{45}
$$ R_{\tau, V+A} = \frac{B(\tau \to \nu_{\tau} +
hadrons_{S=0})}{B(\tau \to \nu_{\tau} e\bar{\nu}_e)}=$$
\be
 = 6
\vert V_{ud} \vert^2 S_{EW}~ \int\limits^{m^2_{\tau}}_{0}~
\frac{ds}{m^2_{\tau}} \Biggl ( 1 - \frac{s}{m^2_{\tau}} \Biggr )^2
\Biggl [ \Biggl ( 1 + 2 \frac{s}{m^2_{\tau}} \Biggr ) (v_1 + a_1
+a_0)(s) - 2 \frac{s}{m^2_{\tau}} a_0(s) \Biggr ] \label{48}\ee
where $\vert V_{ud} \vert = 0.9735 \pm 0.0008$ is the matrix
element of the Kobayashi-Maskawa matrix, $S_{EW} = 1.0194 \pm
0.0040$ is the electroweak correction \cite{46}. Only one-pion
state is practically contributing to the last term in \cite{48}
and it appears to be small:
\be
\Delta R^{(0)}_{\tau} = - 24 \pi^2 \frac{f^2_{\pi}
m^2_{\pi}}{m^4_{\tau}} = - 0.008 \label{49}\ee Denote
\be
\omega(s) \equiv v_1 + a_1 +a_0 = 2\pi Im [\Pi^{(1)}_V(s) +
\Pi^{(1)}_A(s) + \Pi^{(0)}_A(s) ] \equiv 2 \pi Im \Pi(s)\label{50}
\ee As follows from eq.(\ref{47}), $\Pi(s)$ has no kinematical
pole, but only right-hand cut. It is convenient to transform the
integral in eq.(\ref{48}) into that over the circle of radius
$m^2_{\tau}$ in the complex $s$ plane \cite{47}-\cite{49}:
\be
R_{\tau,\, V+A} = 6\pi i |V_{ud}|^2 S_{EW}
\oint_{|s|=m_\tau^2}\!{ds\over m_\tau^2} \left( 1-{s\over
m_\tau^2} \right)^2 \left( 1+2 {s\over m_\tau^2}\right) \Pi (s) +
\Delta R_\tau^{(0)} \label{51} \ee Eq.(\ref{51}) allows one to
express $R_{\tau,V+A}$ in  terms of $\Pi(s)$ at large $\mid
s\mid=m^2_{\tau}$, where  perturbative theory and OPE are valid.

Calculate first the perturbative contribution into eq.(\ref{51}).
To this end, use the Adler function $D(Q^2)$:
\be
D(Q^2) \, \equiv \, - 2\pi^2 \,{ d\Pi(Q^2)\over d\ln{Q^2}}
\,=\,\sum_{n\ge 0} K_n a^n
 \; , \qquad a\equiv {\alpha_s \over \pi}\; , \qquad  Q^2\equiv
 -s,\label{52} \ee
the perturbative expansion of which is known up to terms $\sim
\alpha^4_s$. In $\overline{MS}$ regularization scheme $K_0 = K_1 =
1$,~~ $K_2 = 1.64$ \cite{50}, $K_3 = 6.37$ \cite{51} for 3
flavours and for $K_4$ there are the estimates $K_4 = 25 \pm 25$
\cite{52} and $K_4=27\pm 16$ \cite{53}. The renormgroup equation
yields
\be
{d a \over d \ln{Q^2}} \, =\, -\beta(a) \,=\, - \sum_{n\ge 0}
\beta_n a^{n+2} \label{53}\ee
 \be \ln{Q^2\over \mu^2}\, = \,-\, \int_{a(\mu^2)}^{a(Q^2)}
 {da\over \beta(a)},
\label{54} \ee in the $\overline{MS}$ scheme for three flavours
$\beta_0 = 9/4$,~$\beta_1 = 4$, ~ $\beta_2 = 10.06$, ~ $\beta_3 =
47.23$ ~ \cite{54,55,Czakon}. Integrating over eq.(\ref{52}) and
using eq.(\ref{53}) we get
\be
\Pi(Q^2)\,=\,{1\over 2\pi^2} \int_{a(\mu^2)}^{a(Q^2)} D(a)
{da\over \beta(a)} \label{55} \ee

Put $\mu^2 = m^2_{\tau}$ and choose some (arbitrary) value
$a(m^2_{\tau})$. With the help of eq.(\ref{54})  one may determine
then $a(Q^2)$ for any $Q^2$ and by analytical continuation for any
$s$ in the complex plane. Then, calculating (\ref{55}) find
$\Pi(s)$ in the whole complex plane. Substitution of $\Pi(s)$ into
eq.(\ref{51}) determines $R_{\tau}$ for the given $a(m^2_{\tau})$
up to power corrections. Thereby, knowing $R_\tau$ from experiment
it is possible to find the corresponding to it $a(m^2_\tau)$.
Note, that with such an approach there is no need to expand the
denominator in eqs.(\ref{54}),(\ref{55}) in the inverse powers of
$ln Q^2/\mu^2$.  Advantages  of transformation of the integral
over the real axis (\ref{48}) in the contour integral are the
following. It can be expected that the applicability region of the
theory presented as perturbation theory (PT) + operator product
expansion (OPE) in the complex $s$-plane is off the dashed region
in Fig.1. It is evident that at positive and comparatively small
$s$ PT+OPE does not work.


\begin{figure}[tb]
\hspace{40mm} \epsfig{file=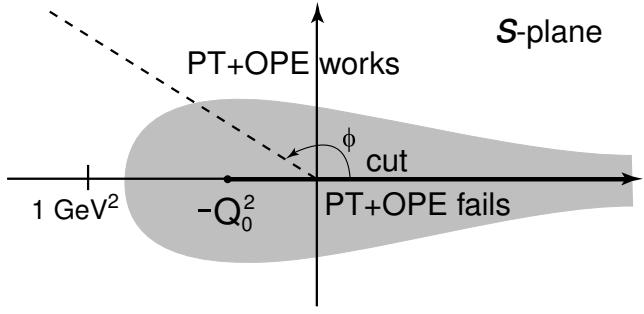, width=85mm} \caption{ The
applicability region of PT and OPE in the complex plane $s$. In
the dashed region PT + OPE does not work.}
\end{figure}

As is well known, in perturbation theory, in the expansion over
the powers of inverse $ln Q^2$, in the first order in $1/ln Q^2$
the running coupling constant $\alpha_s(Q^2)$ has an unphysical
pole at some $Q^2=Q^2_0$. If $\beta(a)$ is kept in the denominator
in (\ref{54}), then in $n$-loop approximation $(n > 1)$ a branch
cut with a singularity $\sim (1-Q^2/Q^2_0)^{-1/n}$ appears instead
of pole. The position of the singularity is given by \be
ln\frac{Q^2_0}{\mu^2} =
-\int\limits^{\infty}_{a(\mu^2)}\frac{da}{\beta(a)}\label{56}\ee
Near the singularity the last term in the expansion of $\beta(a)$
 dominates  and gives  the beforementioned behavior. Since the
singularity became weaker, one may expect a better convergence of
series, which would allow one to go to lower $Q^2$.

\begin{figure}[tb]
\hspace{30mm} \epsfig{file=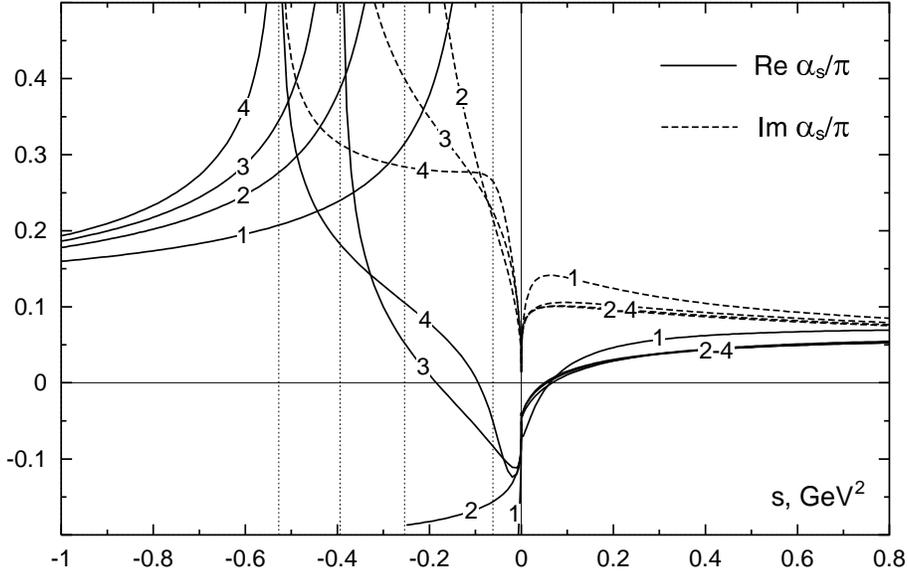, width=120mm} \caption{Real
and imaginary parts of $\alpha_{\overline{\rm MS}}(s)/\pi$  as an
exact numerical solution of RG equation (\ref{54}) on real axes
for different number of loops. The initial condition is chosen
$\alpha_s=0.355$ at $s=-m_\tau^2$, $N_f=3$. Vertical dotted lines
display the position of the unphysical singularity at $s=-Q_0^2$
for each approximation ($4\to 1$ from left to right).}
\label{alpha}
\end{figure}

The real and imaginary parts of $\alpha_s(s)/\pi$, obtained as
numerical solutions of eq.(\ref{54}) for various numbers of loops
are plotted in Fig.2 as  functions of $s=-Q^2$. The $\tau$-lepton
mass was chosen as normalization point, $\mu^2=m^2_{\tau}$ and
$\alpha_s(m^2_{\tau})=0.355$ was put in. As is  seen from Fig.2,
at negative $s$ the perturbation theory  converges at $s < -1
GeV^2$ and in order to have a good precision of the results 4
loops calculations are necessary. At positive $s$, especially for
$Im(\alpha_s/\pi)$, the convergence of the series is much better.
This comes from the fact, that in the chosen integral form of
renormalization group equation (\ref{54}) the expansion over
$\pi/ln(Q^2/\Lambda^2)$ is avoided, this expansion  being not  a
small parameter at intermediate $Q^2$. (The systematical method of
analytical continuation from the spacelike to timelike region with
summation of $\pi^2$ terms was suggested in \cite{56} and
developed in \cite{57}). For instance, in the next to leading
order \be 2\pi Im\Pi(s+i0) = 1+ \frac{1}{\pi \beta_0} \Biggl [
\frac{\pi}{2} -arctg \Biggl ( \frac{1}{\pi} ln
\frac{s}{\Lambda^2}\Biggr ) \Biggr ]\label{57}\ee instead of
\be
2\pi Im\Pi (s+i0)=1 +\frac{1}{\beta_0
ln(s/\Lambda^2)},\label{58}\ee which would follow in the case of
small $\pi/ln(s/\Lambda^2)$.

The $\alpha_s(Q^2)$ at $Q^2 > 0$ in low $Q^2$ region $(0.8 < Q^2 <
5 GeV^2)$ is plotted in Fig.3. (Four loops are accounted,
$\alpha_s(m^2_{\tau})$ is put to be equal to
$\alpha_s(m^2_{\tau})=0.33$. As follows from  $\tau$-decay rate
$\alpha_s(m^2_{\tau})=0.352\pm 0.020$ and the value of one
standard deviation below the mean one is favoured by low energy
sum rules).

 Integration over the contour allows one to
obviate the dashed region in Fig.1 (except for the vicinity of the
positive semiaxis, the contribution of which is suppressed  by the
factor $(1 - \frac{s}{m^2_{\tau}})^2$ in eq.(\ref{51})), i.e. to
work in the applicability region of PT+OPE.


\begin{figure}[tb]
\hspace{40mm} \epsfig{file=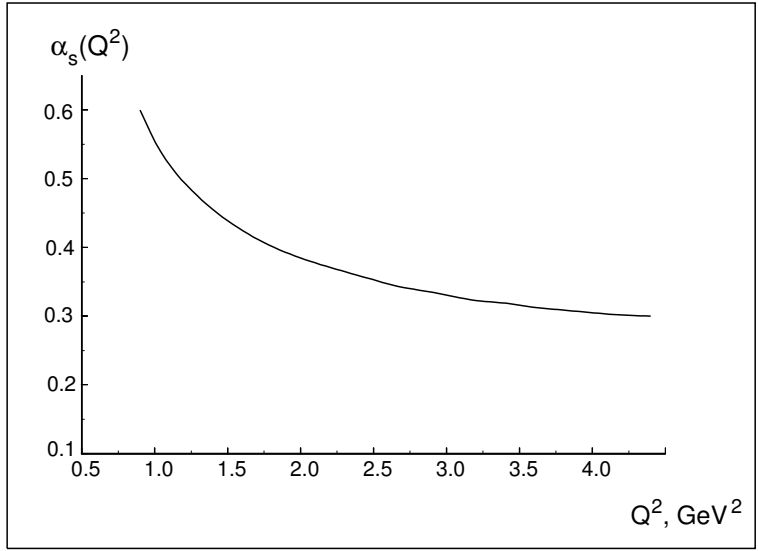, width=100mm}
\caption{$\alpha_s(Q^2)$ normalized at $m_{\tau}^2$,
$\alpha_s(m^2_{\tau})=0.33$.}
\end{figure}

The OPE terms, i.e., power corrections to polarization operator,
are given by the formula \cite{1}:
\bea \Pi(s)_{nonpert} &=&\sum_{n\ge 2} {\left<O_{2n}\right>\over
(-s)^n} \left( 1+ c_n {\alpha_s\over \pi} \right)\nonumber \\ & =
&
 {\alpha_s\over 6 \pi\, Q^4} \left< 0
 \mid G_{\mu\nu}^aG_{\mu\nu}^a \mid 0 \right>\left( 1
 +
{7\over 6} {\alpha_s\over \pi} \right)+\frac{4}{Q^4}(m_u+m_d)
\left < 0\mid \bar{q}q\mid 0 \right > \nonumber \\ & + & {128\over
81\, Q^6}\, \pi\alpha_s \left<0 \mid \bar{q}q\mid 0
\right>^2_{\mu} \left[ 1 + \left({29\over 24} + {17\over
18}\ln{Q^2\over \mu^2} \right){\alpha_s\over \pi}
 \right] + {\left<O_8\right>\over Q^8}  \label{59}
\eea
 ($\alpha_s$-corrections to the 1-st and 3-d terms in
eq.(\ref{59}) were calculated in \cite{58} and \cite{59},
respectively). Contributions of terms proportional to $m^2_u$,
$m^2_d$ are neglected.  When calculating the d=6 term,
factorization hypothesis was used. Gluon condensate of dimension
$d=6~~g^3\langle 0\mid G^3 \mid 0\rangle$ (\ref{28}) does not
contribute to polarization operator (\ref{59}). This is a
consequence of the general theorem, proved by Dubovikov and Smilga
\cite{DS}, that in case of self-dual gluonic fields there are no
contributions of gluon  condensates of dimensions higher than
$d=4$ to vector and axial currents polarization operators. Since
the  vacuum expectation value of $G^3$ operator does not vanish
for self-dual gluonic fields, this means the vanishing of the
coefficient in front of $g^3 \langle 0\mid G^3 \mid 0\rangle $
condensate in (\ref{59}). The same argument refers to dimension 8
gluon operators $g^4G^4$ with the exception of some of them, like
$g^4[~G^n_{\mu\alpha} G^n_{\mu\beta} - (1/4)
\delta_{\alpha\beta}G^n_{\mu\nu}G^n_{\mu\nu}~]^2$, which have zero
expectation values in any self-dual field. But the latter are
suppressed by a small factor $1/4\pi^2$ arising from loop
integration  in comparison with tree diagram, corresponding to
$d=8$ four quark condensate $\langle O_8\rangle\sim \langle
\bar{q} G q \cdot \bar{q}q\rangle$ contribution. The contribution
from this condensate may be estimated as $\mid \langle O_8\rangle
\mid < 10^{-3}~GeV$ \cite{42} (see below, Sec.5.1) and appears to
be negligibly small. The $d=8$ two quarks -- two gluons operator
$O'_8 \sim g^2 D\bar{q}GG q$ is nonfactorizable,  its vacuum mean
value is suppressed by $1/N_c$ and one may believe, that its
contribution to (\ref{59}) is also small.
 It can be
readily seen that d=4 condensates (up to small $\alpha_s$
corrections) give no contribution into the integral over contour
eq.(\ref{51}). $R_{\tau,V+A}$ may be represented as
\be R_{\tau, V+A}  = 3|V_{ud}|^2 S_{EW}\left(
\,1\,+\,\delta_{em}'\,+\,\delta^{(0)}\,+\,\delta^{(6)}_{V+A} \,
\right) +\Delta R^{(0)}~=~3.486 \pm 0.016 \label{60} \ee
 where
$\delta^{\prime}_{em} = (5/12 \pi)\alpha_{em}(m^2_{\tau}) = 0.001$
is the electromagnetic correction \cite{60}, $\delta^{(6)}_{A+V} =
-(3.3\pm 1.1)\cdot 10^{-3}$ is the contribution of d=6 condensate
(see below) and $\delta^{(0)}$ is the PT correction. The
right-hand part presents the experimental value obtained as a
difference between the total probability of hadronic decays
$R_{\tau} = 3.647 \pm 0.014$ \cite{Davier} and the probability of
decays in states with the strangeness $S = -1 ~~R_{\tau,s} = 0.161
\pm 0.007$ \cite{61,62}.  For perturbative correction it follows
from eq.(\ref{60}), that
\be \delta^{(0)} = 0.208 \pm 0.006 \label{61}\ee From (\ref{61})
employing the above described method,  the constant
$\alpha_s(m^2_{\tau})$ was found  \cite{42}
\be
\alpha_s(m^2_{\tau}) = 0.352 \pm 0.020  \label{62}\ee The
calculation was made with the account of terms $\sim
\alpha^4_{\tau}$, the theoretical error was assumed to be equal to
the last term contribution. May be, the error is underestimated
(by $\sim 0.010$), since the theoretical and experimental errors
were added in quadratures. The value $\alpha_s(m^2_{\tau})$
(\ref{62}) corresponds to: \be \alpha_s(m^2_z)=0.121 \pm
0.002\label{63}\ee

This value is in agreement with recent determination \cite{a1} of
$\alpha_s(m^2_z)$ from the whole set data
\be
\alpha_s(m^2_z)= 0.1182 \pm 0.0027 \label{64a}\ee

\subsection{\it Instanton corrections}

Some nonperturbative features of QCD may be described in the so
called instanton gas model (see \cite{SShur} for extensive review
and the collection of related papers in \cite{S2}). Namely, one
computes the correlators in the $SU(2)$-instanton field embedded
in the $SU(3)$ color group. In particular, the 2-point correlator
of the vector currents had been computed long ago \cite{AG}. Apart
from the usual tree-level correlator $\sim \ln {Q^2}$ it has a
correction which depends on the instanton position and radius
$\rho$. In the instanton gas model these parameters are integrated
out. The radius is averaged over some concentration $n(\rho)$, for
which one or another model is used. Concerning the 2-point
correlator of charged axial currents, the only difference from the
vector case is that the term with 0-modes must be taken with
opposite sign. In coordinate representation the answer can be
expressed in terms of elementary functions, see \cite{AG}. An
attempt to compare the instanton correlators with the ALEPH data
in the coordinate space, was made  in Ref.\cite{SShur2}.

We shall work in momentum space. Here the instanton correction to
the spin-$J$ parts $\Pi^{(J)}$ of the correlator (\ref{45}) can be
written in the following  form: \bea \Pi^{(1)}_{V,\,{\rm
inst}}(q^2)& = & \int_0^\infty\!d\rho \, n(\rho)\, \left[\,
-\,{4\over 3 q^4}\,+\,\sqrt{\pi}\rho^4 G^{30}_{13} \left( -\rho^2
q^2 \left| { 1/2 \atop 0,0,-2 }\right. \right) \right]\nonumber \\
\Pi^{(0)}_{A,\,{\rm inst}}(q^2) & = &\int_0^\infty\!d\rho \,
n(\rho)\, \left[ \,-\,{4\over q^4}\, - \,{4\rho^2\over
q^2}\,K_1^2\!\left(\rho\sqrt{-q^2}\right) \right]\nonumber \\
\Pi^{(1)}_{A,\, {\rm inst}}(q^2) &  =  &\Pi^{(1)}_{V,\, {\rm
inst}}(q^2)-\Pi^{(0)}_{A,\, {\rm inst}}(q^2) \; , \qquad
\Pi^{(0)}_{V,\, {\rm inst}}(q^2)\,=\,0 \label{64} \eea Here $K_1$
is modified Bessel function, $G_{mn}^{\,p\,q}(z|\ldots)$ is Meijer
function. Definitions, properties and approximations of Meijer
functions can be found, for instance, in \cite{Luke}. In
particular the function in (\ref{64}) can be written as the
following series:  \bea \sqrt{\pi}G^{30}_{13}\left( z\left| { 1/2
\atop 0,0,-2 }\right. \right)& = &{4\over 3z^2}\,-\,{2\over z}\,
+\,{1\over 2\sqrt{\pi}}\,\sum_{k=0}^\infty \,z^k
{\Gamma(k+1/2)\over \Gamma^2(k+1)\,\Gamma(k+3)}\nonumber
\\ &\times &
\Bigl\{\, \left[\, \ln{z}\,+\,\psi(k+1/2)\,-\,2\,\psi(k+1)\,  -
\,\psi(k+3)\, \right]^2 \nonumber \\ & + &
\,\psi'(k+1/2)\,-\,2\,\psi'(k+1)\,-\,\psi'(k+3) \,\Bigr\}
\label{65} \eea where  $\psi(z)=\Gamma'(z)/\Gamma(z)$. For large
$|z|$ one can obtain its approximation by the saddle-point method:
\be G^{30}_{13}\left( z\left| { 1/2 \atop 0,0,-2 }\right. \right)
\approx \sqrt{\pi} z^{-3/2} e^{-2\sqrt{z}} \; , \qquad |z|\gg 1
\label{66} \ee

The formulae (\ref{64}) should be treated in the following way.
One adds $\Pi_{\rm inst}$ to usual polarization operator  with
perturbative and OPE terms. But the terms $\sim 1/q^4$
\underline{must be absorbed} by the operator $O_4$ in
Eq.(\ref{64}), since the gluonic condensate $\bigl<G^2\bigr>$ is
averaged over all field configurations, including the instanton
one. Notice negative sign before $1/q^4$ in Eq.(\ref{64a}). This
happens because the negative contribution of the quark condensate
$\bigl<m\bar{q}q\bigr>$ in the instanton field exceeds positive
contribution of the gluonic condensate $\bigl<G^2\bigr>$. In real
world $\bigl<m\bar{q}q\bigr>$ is negligible.

The correlators (\ref{64}) possess appropriate analytical
properties, they have a cut along positive real axes: \be {\rm
Im}\,\Pi^{(1)}_{V,\, {\rm inst}}(q^2+i0)  = \int_0^\infty\!d\rho
\, n(\rho)\,\pi^{3/2} \rho^4 G^{20}_{13} \left(\rho^2 q^2 \left| {
1/2 \atop 0,0,-2 }\right. \right) \label{67} \ee
\be
 {\rm Im}\,\Pi^{(0)}_{A,\, {\rm inst}}(q^2+i0) = -\,\int_0^\infty\!d\rho \, n(\rho)\,
{2\pi^2\rho^2\over q^2}
J_1\!\left(\rho\sqrt{q^2}\right)N_1\!\left(\rho\sqrt{q^2}\right)
\label{68} \ee

We shall consider below the instanton gas model.  It is a model
with fixed instanton radius
\be
n(\rho)\,=\,n_0 \, \delta(\rho - \rho_0)\label{69} \ee In
\cite{SShur} it was  estimated:
\be
\rho_0\,\approx \,1/3\,{\rm fm} \, \approx \,1.5 - 2.0 \, {\rm
GeV}^{-1} \; , \qquad n_0\,\approx \,1\,{\rm fm}^{-4} \, \approx
(1.0 - \,1.5) \times 10^{-3} \, {\rm GeV}^4 \label{70} \ee In
fact, the instanton liquid model, with the account of instanton
self-interaction was mainly considered in \cite{SShur}, but the
arguments, from which the estimations (\ref{70}) follow, refer
also to the instanton gas model. In this case, the value of $n_0$
(\ref{70}) should be considered as an upper limit (see also
\cite{EVShuryak}).

\begin{figure}[tb]
\hspace{10mm}\epsfig{file=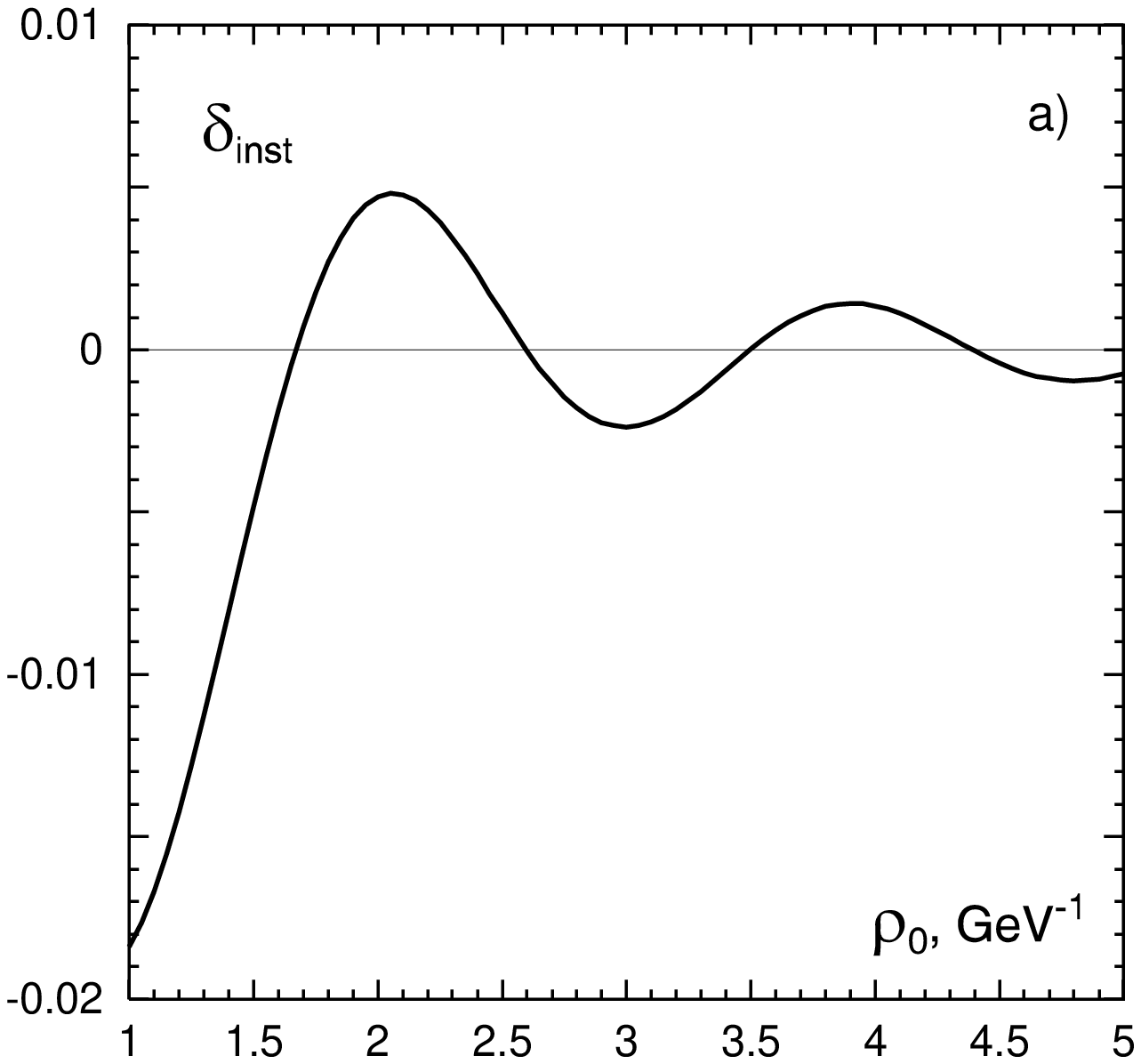, width=81mm}
\epsfig{file=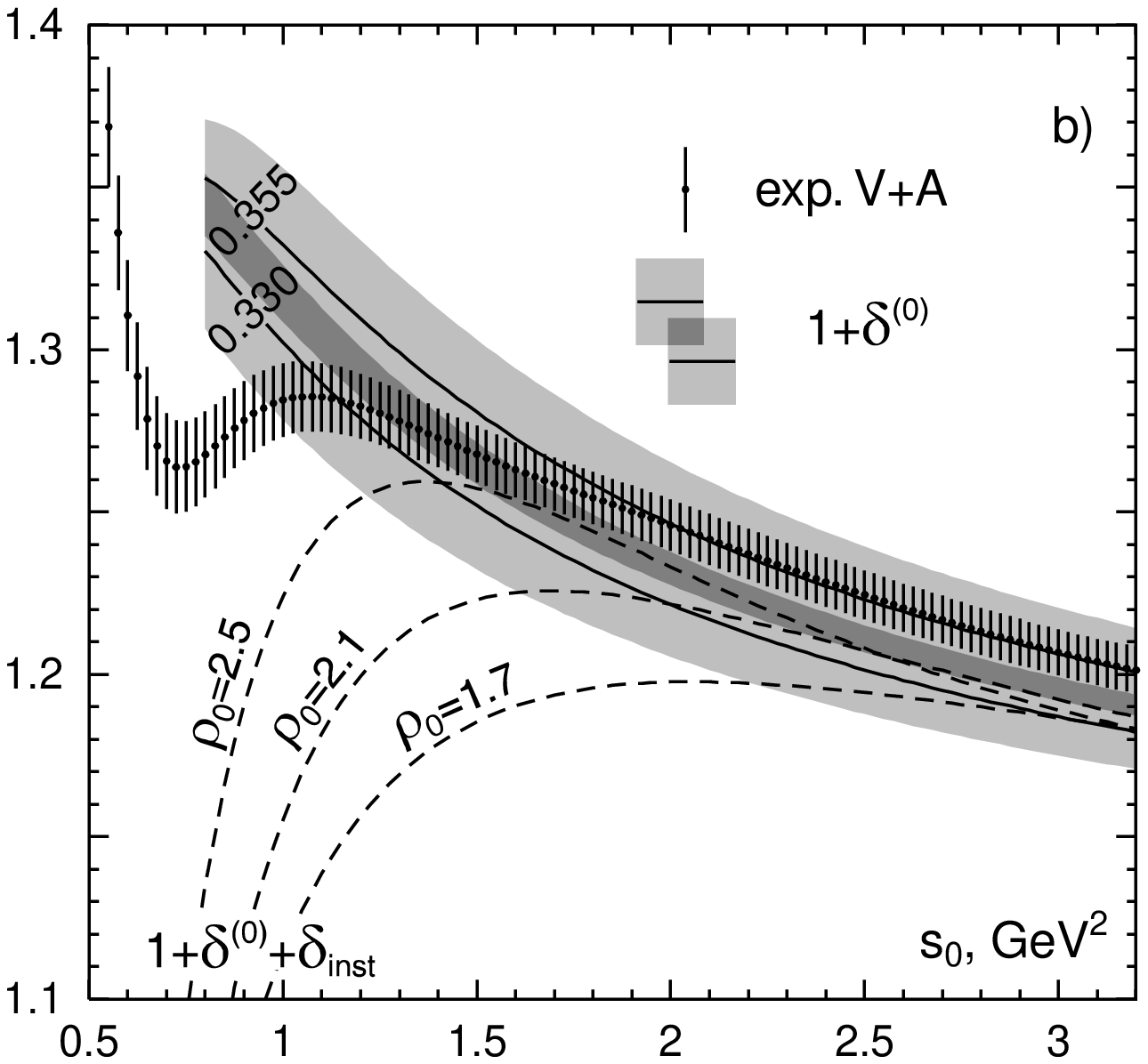, width=81mm} \caption{The instanton
correction to the $\tau$ decay ratio versus $\rho_0$ (a) and
 "versus $\tau$ mass" (b) for $n_0=1.5 \times 10^{-3}\,{\rm GeV}^4$.
 The thin solid lines in Fig.\ref{inst_tau}b are the values of
 $1+\delta^{(0)}(s_0)$, where $\delta^{(0)}(s_0)$ are
 perturbative corrections, calculated as described in Sec.4.1.
 The upper curve corresponds to $\alpha_s(m^2_{\tau})=0.355$, the
 lower one -- to $\alpha_s(m^2_{\tau})=0.330$. The shadowed regions
 represent the uncertainties in perturbative calculations, the
 dark shadowed band is their overlop. The dashed lines are
 $1+\delta^{(0)}(s) +\delta_{inst}, \delta^{(0)}(s)$ corresponds
 to $\alpha_s(m^2_{\tau})=0.330$. }
\label{inst_tau}
\end{figure}

Now we consider the instanton contribution to the $\tau$-decay
branching ratio. Since the instanton correlator (\ref{64}) has
$1/q^2$ singular term in the expansion near 0 (see Eq.
(\ref{65})), the integrals must be taken over the circle, as in
(\ref{51}). In the instanton model the function $a_0(s)$ differs
from experimental $\delta$-function, which gives small correction.
So we shall ignore the last term in (\ref{48}) and consider the
integral with $\Pi_{V+A}^{(1)}+\Pi_A^{(0)}$ in (\ref{51}). The
instanton correction to the $\tau$-decay branching ratio can be
brought to the following form:
\be
\delta_{\rm inst}\,  =\,  -\,48\,\pi^{5/2}\int_0^\infty\!d\rho \,
n(\rho)\,\rho^4\, G^{20}_{13}\left(\rho^2 m_\tau^2 \left| { 1/2
\atop 0,-1,-4 }\right. \right) \approx \, {48\pi^2 n_0 \over
\rho_0^2 m_\tau^6} \, \sin{(2\rho_0 m_\tau)} \label{71} \ee Since
the parameters (\ref{70}) are determined quite approximately, we
may explore the dependence of $\delta_{\rm inst}$ on them. The
$\delta_{\rm inst}$ versus $\rho_0$ for fixed $n_0=1.5\cdot
10^{-3} GeV^4$  is shown in Fig.\ref{inst_tau}.

As seen from Fig.\ref{inst_tau}a the instanton correction to
hadronic $\tau$-decay is extremely small except for unreliably low
value of the instanton radius $\rho_0<1.5\,{\rm GeV}^{-1}$. At the
favorable value \cite{SShur} $\rho_0=1.7\,{\rm GeV}^{-1}$ the
instanton correction to $R_\tau$ is almost exactly zero. This fact
confirms the calculations of $\alpha_s(m_\tau^2)$ (Sec.~4.1),
where the instanton corrections were not taken into account.

Eq.(\ref{71}) can be used in another way. Namely, the $\tau$ mass
can be considered as free parameter $s_0$. The dependence of the
fractional corrections $\delta^{(0)}$ and
$\delta^{(0)}_{0.330}+\delta_{\rm inst}$ on $s_0$ is shown in
Fig.\ref{inst_tau}b.  The result strongly depends on the instanton
radius and rather essentially on the density $n_0$. For
$\rho_0=1.7\,{\rm GeV}^{-1}$ and $n_0=1\,{\rm fm}^{-4}$,  the
instanton curve is outside the errors already at $s_0\sim 2\,{\rm
GeV}^2$, where the perturbation theory is expected to work.

We came to the conclusion, that  in case of variable $\tau$ mass
the instanton contribution becomes large at $s_0 < 2 GeV^2$. That
means, that $R_{\tau,V+A}(s_0)$ given by (\ref{51}) cannot be
represented by PT+OPE at $s_0 < 2 GeV^2$ and the results, obtained
in this way are not reliable.

\subsection{\it Comparison with other approaches}

There are many calculations of $\alpha_s(m^2_{\tau})$ from the
total $\tau$-decay rate, using the same idea, which was used above
-- the countour improved fixed order perturbation  theory
\cite{43,45,47}-\cite{49,Davier}. (For more recent ones, see
\cite{15,44}.) The results of these calculations coincide with
presented above in the limit of errors and give
$\alpha_s(m^2_{\tau})=0.33-0.35$. From these values by using
renormalization  group one can find $\alpha_s(m^2_z)=0.118-0.121$
in agreement with $\alpha_s(m^2_z)$ determinations from other
processes (see \cite{10},\cite{a1}).

Till now only one renormalization scheme was considered -- the
$\overline{MS}$ scheme. In BLM renormalization scheme \cite{68},
which have some advantages from the point of view of perturbative
pomeron theory \cite{69}, the result is
$\alpha_s(m^2_{\tau})=0.621\pm 0.008$ \cite{70}, corresponding in
the framework of BLM scheme to the same value of
$\alpha_s(m^2_z)=0.117-0.122$. At low scales, however, the
$\alpha_s(Q^2)$ behavior is essentially different from that,
presented in Fig.3.

Few words about $\alpha_s$ calculations in analytical QCD (see
\cite{71} and references herein). According to this theory the
coupling constant $\alpha_s(Q^2)$ is calculated by renormalization
group in the spacelike region $Q^2 >0$. Then, by analytical
continuation to $s=-Q^2 >0$ $Im \alpha_s(s)$ was found on the
right semiaxes. It was assumed, that $\alpha_s(s)$ is an
analytical function in the complex $s$-plane with a cut along the
right semiaxes $0\leq s\leq\infty$. The analytical
$\alpha_s(s)_{an}$ is then defined in the whole $s$-plane by
dispersion relation. Such $\alpha_s(s)_{an}$ has no unphysical
singularities. Let us calculate $\alpha_s(m^2_{\tau})_{an}$ using
the same experimental data as before, i.e. $\delta^{(0)}$ given by
Eq.(\ref{61}). In the analytical QCD the countour  integral
(\ref{51}) is equal to the integral of $Im \Pi(s)$ over real
positive axes. (In the previous calculation the integral was
running  from $s=-Q^2_0$ to $m^2_{\tau}$.) Qualitatively, it leads
to much smaller $R_{\tau}$ in the analytical QCD than in the
conventional  approach with the same $\alpha_s(m^2_{\tau})$, or
vice versa, it is necessary to  have much larger
$\alpha_s(m^2_{\tau})_{an}$ in order to get experimental
$R_{\tau}$. The calculation  of integral (\ref{51}) with $\Pi(s)$
expressed through $\alpha_s(s)_{anal}$, shows that experimental
$R_{\tau}$ results in $\alpha_s(m^2_z)=0.141\pm 0.004$ in
contradiction with other data.  (In \cite{72} an attempt was made
to get an agreement of analytical QCD with common value of
$\alpha_s(m^2_z)$. For this goal the  constituent quark model with
specific quark-antiquark potential was used in the domain of low
and intermediate $s$. Evidently, such approach cannot be
considered as $\alpha_s$ determination in QCD: in this approach
QCD is modified on large circle in complex plane of the radius
$\mid s\mid =m^2_{\tau}$ in contradiction with the basic
assumption of $\alpha_s$ calculation from hadronic $\tau$-decay
rate.)

\bigskip

\section{Determination of condensates from spectral functions of
$\tau$-decay}

\subsection{\it Determination of quark condensate  from $V-A$ spectral
function}

In order to determine  the quark condensate from $\tau$-decay data
it is convenient to consider the difference $V-A$ of polarization
operators $\Pi^{(1)}_V - \Pi^{(1)}_A$, where the contribution of
perturbative  terms is absent. $\Pi^{(1)}_V(s)-\Pi^{(1)}_A(s)$ is
represented by OPE:
\be
\Pi^{(1)}_V(s) -\Pi^{(1)}_A(s) =\sum_{D\geq 4}
\frac{O^{V-A}_D}{(-s)^{D/2}} \Biggl (1+
c_D\frac{\alpha_s(s)}{\pi}\Biggr )\label{73}\ee The gluonic
condensates contribution drops out in the $V-A$ difference and
only the following condensates up to D=10 remain
\be O^{V-A}_4  =  2 \,(m_u +m_d)\,\langle 0 \mid \bar{q}q \mid 0
\rangle \; = \; -\, f_\pi^2 m_\pi^2~[1] \label{74}\ee
$$ O^{V-A}_6  =  2\pi \alpha_s \left< 0 \mid
(\bar{u}\gamma_\mu\lambda^a d)(\bar{d}\gamma_\mu \lambda^a u) -
(\bar{u}\gamma_5\gamma_\mu\lambda^a d)(\bar{d}\gamma_5\gamma_\mu
\lambda^a u)\mid 0 \right>  = $$ \be = -\,{64\pi\alpha_s\over 9}
\langle 0\mid \bar{q}q\mid 0 \rangle^2~[1] \label{75} \ee
\be
 O^{V-A}_8  =   -8\pi \alpha_s \, m_0^2
\langle 0 \mid \bar{q}q \mid
0\rangle^2~,~\cite{Dubovikov,41}~\footnote{There was a sign error
in the contribution of $O_8$ in \cite{41}.} \label{76} \ee
\be
O^{V-A}_{10} =-\frac{8}{9} \pi\alpha_s \langle 0 \mid \bar{q}q\mid
0\rangle^2 \Biggl [\frac{50}{9} m^4_0 +32 \pi^2 \langle 0\mid
\frac{\alpha_s}{\pi} G^2 \mid 0 \rangle \Biggr
]~\cite{76}\label{77}\ee where $m^2_0$ is determined in
eq.(\ref{24}). In the right-hand of
(\ref{75}),(\ref{76}),(\ref{77}) the factorization hypothesis was
used. For $O_6$ operator it is expected \cite{1}, that the
accuracy of factorization hypothesis is of order $1/N^2_c\sim
10\%$, where $N_c=3$ is the number of colours. For operators of
dimensions $d \geq 8$ the factorization procedure is not unique.
(But, as a rule, the arising differences are not very large -- for
$d = 8$  operator entering eq.(\ref{73}) it is about 20\%). The
accuracy of factorization
 hypothesis becomes worse with increasing of
operator dimensions: for $O^{V-A}_8$, it is worse, than for
$O^{V-A}_6$ and for $O^{V-A}_{10}$ it is worse than for
$O^{V-A}_8$.

Operators $O_4$ and $O_6$ have approximately  zero anomalous
dimensions, the $O_8$ anomalous dimension is equal to -- 11/27.
Calculations of the coefficients in front of $\alpha_s$ in
eq.(\ref{73}) gave $c_4=4/3$ \cite{77} and $c_6=89/48$ \cite{78}.
(For $O_4$ the $\alpha^2_s$ correction is known \cite{77}:
$(59/6)(\alpha_s/\pi)^2)$.) The $\alpha_s$ corrections to
$O^{V-A}_8$ are unknown -- they are included  into the not
certainly  known value of $m^2_0$, $\alpha_s$ corrections to
$O_{10}$ are unknown also. (In this Section indeces  $V-A$ will be
omitted and $O_D$ will mean condensates with $\alpha_s$
corrections included.)

Our aim is to compare the OPE theoretical predictions  with the
experimental data on $V-A$ structure functions measured in
$\tau$-decay and with  the help of such comparison  to  determine
the magnitude of the most important condensate $O_6$. The
condensate  $O_4$ is small and is known with a good accuracy:
\be O_4 =-0.5\cdot 10^{-3}~ GeV^4\label{78}\ee We put $m^2_0=0.8
GeV^2$ and in the analysis of the data the values of the
condensates $O_8$ and $O_{10}$ are  taken to be equal to
\be
O_8=-2.8\cdot 10^{-3}~ GeV^8\label{79}\ee
\be O_{10}=-2.6\cdot 10^{-3}~ GeV^{10}\label{80}\ee and their
$Q^2$-dependence, arising from anomalous dimensions is neglected.

In ~the~ calculation ~of ~numerical ~values~ (\ref{78}),(\ref{79})
it~ was ~assumed, ~that ~$a_{\bar{q}q}(1~ GeV^2)\equiv$\\ $\equiv
-(2\pi)^2 \langle 0\mid \bar{q}q\mid 0 \rangle_{1 GeV} = 0.65~
GeV^3$, $\langle 0\mid (\alpha_s/\pi) G^2 \mid 0 \rangle =0.005~
GeV^4$ -- see below, eq.'s (\ref{86}),(\ref{116}).

 As was shown in \cite{41} the dimension $d=8$ four-quark operators
 for vector and axial currents are of opposite sign and equal in
 absolute values up to terms of order $1/N^2_c$: $O^V_8=- O^A_8
 (1+O(N^{-2}_c))$. (The exact value of $N^{-2}_c$ correction is
 uncertain -- it depends on factorization procedure.) So, for
 $O^{V+A}_8$ we have from (\ref{79}) the estimation: $\mid
 O^{V+A}_8 \mid < 10^{-3}~GeV^8$, which was used in calculation
 $\Pi(s)_{nonpert}$, Eq.(\ref{59}).

For $\Pi^{(1)}_V(s)-\Pi^{(1)}_A(s)$  substractionless dispersion
relation is valid:
\be
\Pi^{(1)}_V(s)-\Pi^{(1)}_A(s)\,=\,{1\over 2\pi^2}\int_0^\infty
{v_1(t)-a_1(t)\over t-s} \, dt\,+ \,{f_\pi^2\over s} \label{81}
\ee (The last term in the right-hand part is the kinematic pole
contribution). The experimental data for $v_1(s) - a_1(s)$ are
presented in Fig.5


\begin{figure}[tb]
\hspace{10mm}
\epsfig{file=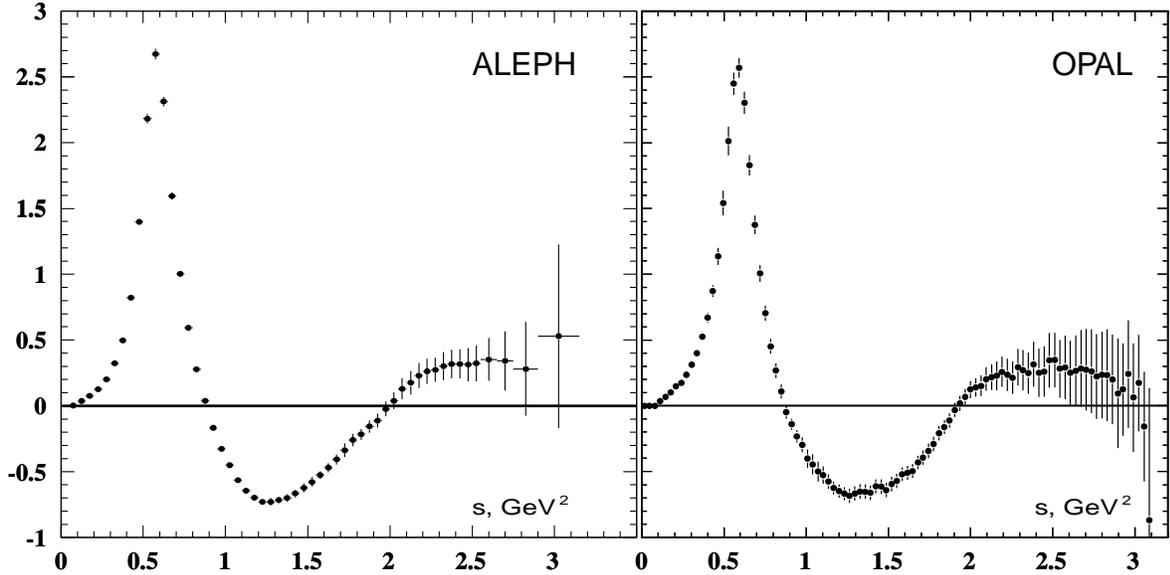}
\caption{The measured difference $v_1(s)-a_1(s)$. Figures from
\cite{38} and \cite{39}, reproduced in \cite{41}.
 \label{vma_exp}}
\end{figure}

In order to improve  the convergence of OPE series as well as to
suppress the contribution of large $s$ domain in dispersion
integral use the Borel transformation.
 Put $s = s_0e^{i \phi}$  ($\phi = 0$ on the upper
edge of the cut) and make the Borel transformation in $s_0$. As a
result, we get the following sum rules for the real and imaginary
parts of (\ref{81}):
\be
\int_0^\infty \exp{\!\left({s\over
M^2}\cos{\phi}\right)}\cos{\!\left({s\over M^2}\sin{\phi}\right)}
(v_1-a_1)(s)\,{ds\over 2\pi^2} \, = \, f_\pi^2+\,\sum_{k=1}^\infty
(-1)^k {\cos{(k\phi)} \,O_{2k+2}\over k!\, M^{2k}} \label{82}\ee
\be
\int_0^\infty \exp{\!\left({s\over
M^2}\cos{\phi}\right)}\,\sin{\!\left({s\over
M^2}\sin{\phi}\right)} (v_1-a_1)(s)\,{ds\over 2\pi^2 M^2} \, =
\,\sum_{k=1}^\infty (-1)^k {\sin{(k\phi)} \,O_{2k+2}\over k!\,
M^{2k+2}} \label{83}\ee
 The use of the Borel transformation along
the rays in the complex plane has a number of advantages. The
exponent index is negative at $\pi/2 < \phi < 3 \pi/2$. Choose
$\phi$ in the region $\pi/2 < \phi < \pi$. In this region, on one
hand, the shadowed area in Fig.~1 in the integrals
(\ref{82}),(\ref{83}) is touched to a less degree, and on the
other hand, the contribution of large $s$, particularly, $s >
m^2_{\tau}$ , where experimental data are absent, is exponentially
suppressed. At definite values of $\phi$ the contribution of some
condensates vanishes, what may be also used.  In particular, the
condensate $O_8$ does not contribute to (\ref{82}) at $\phi = 5
\pi/6$ and to (\ref{83}) at $\phi = 2 \pi/3$, while  the
contribution of $O_6$ to (\ref{82}) vanishes at $\phi = 3 \pi/4$.
Finally, a well known advantage of the Borel sum rules is
factorial suppression of higher dimension terms of OPE. Figs.6,7
present the results of the calculations of the left-hand parts of
eqs.(\ref{82}),(\ref{83}) on the basis of the ALEPH \cite{38}
experimental data comparing with OPE predictions -- the right-hand
part of these equations.


\begin{figure}[tb]
\hspace{20mm} \epsfig{file=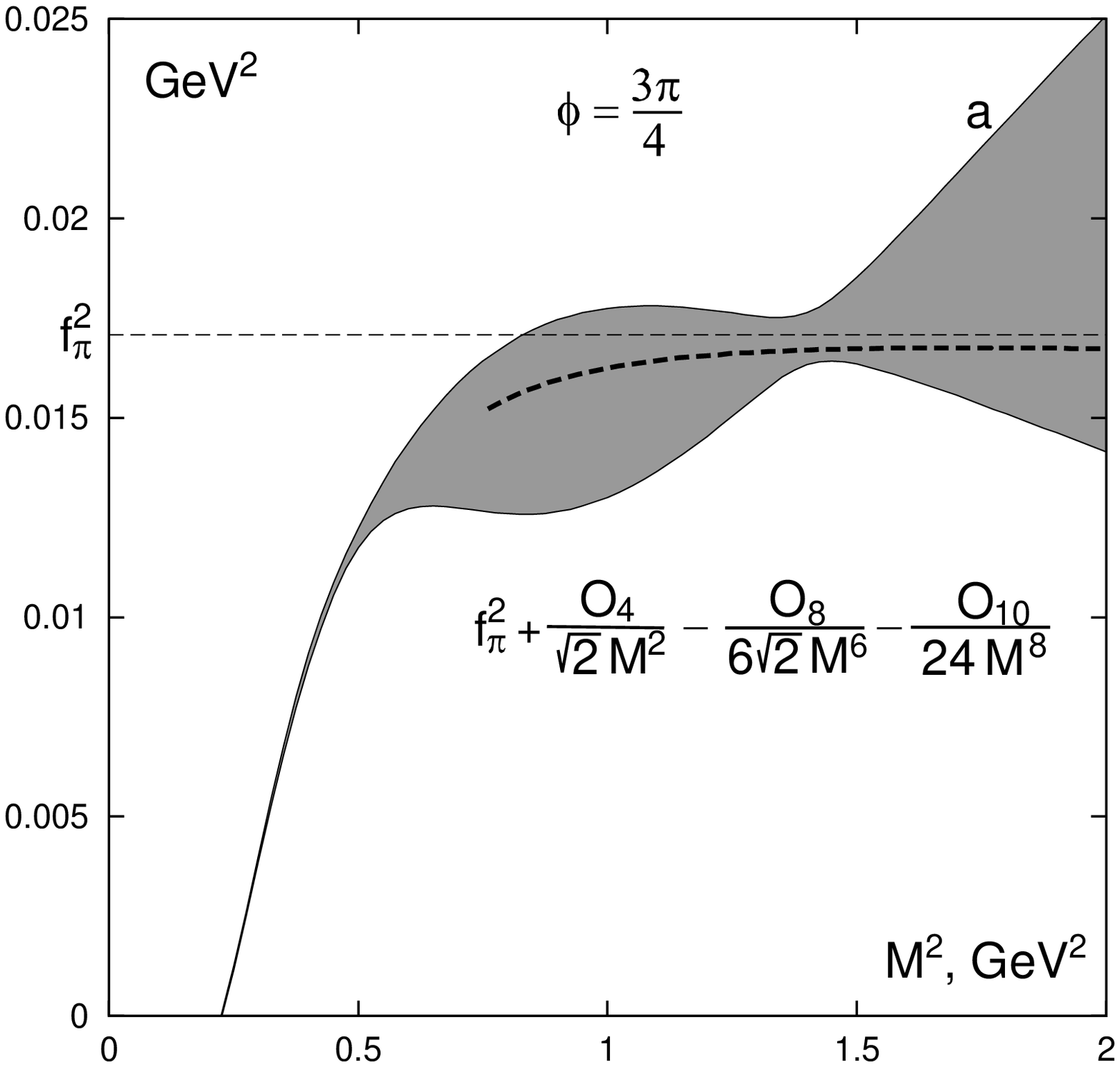, width=70mm} \hspace{5mm}
\epsfig{file=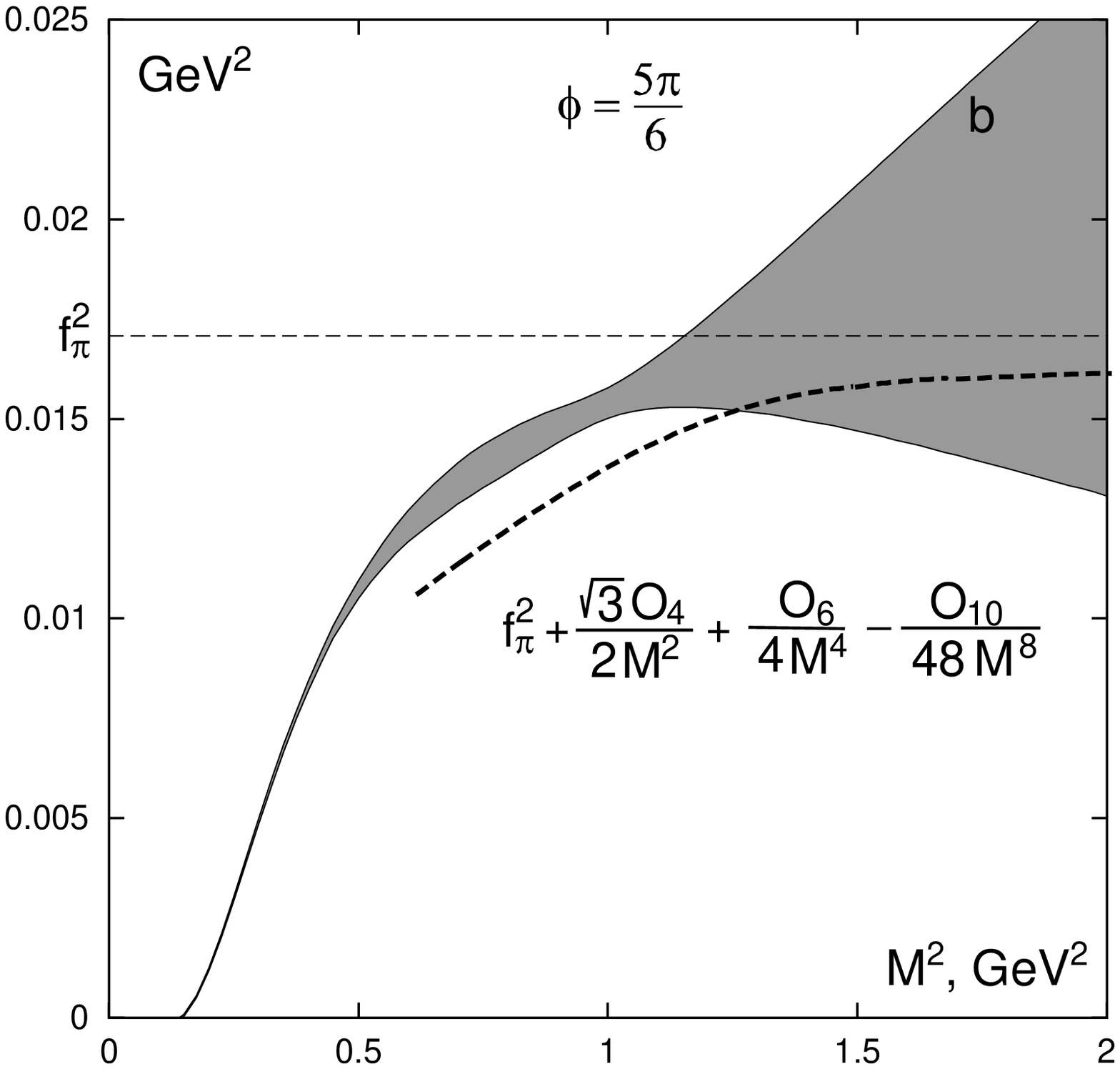, width=70mm} \caption{Eq.(\ref{82}): the
left-hand part is obtained basing on the experimental data, the
shaded region corresponds to experimental errors; the right-hand
part -- the theoretical one -- is represented by the dotted curve,
numerical values of condensates  $O_4,O_8,O_{10},O_6$ are taken
according to (\ref{78}),(\ref{79}),(\ref{80}),(\ref{84}); a) $\phi
= 3 \pi/4$,~ b) $\phi = 5 \pi/6$.}
\end{figure}
\begin{figure}[tb]
\hspace{20mm} \epsfig{file=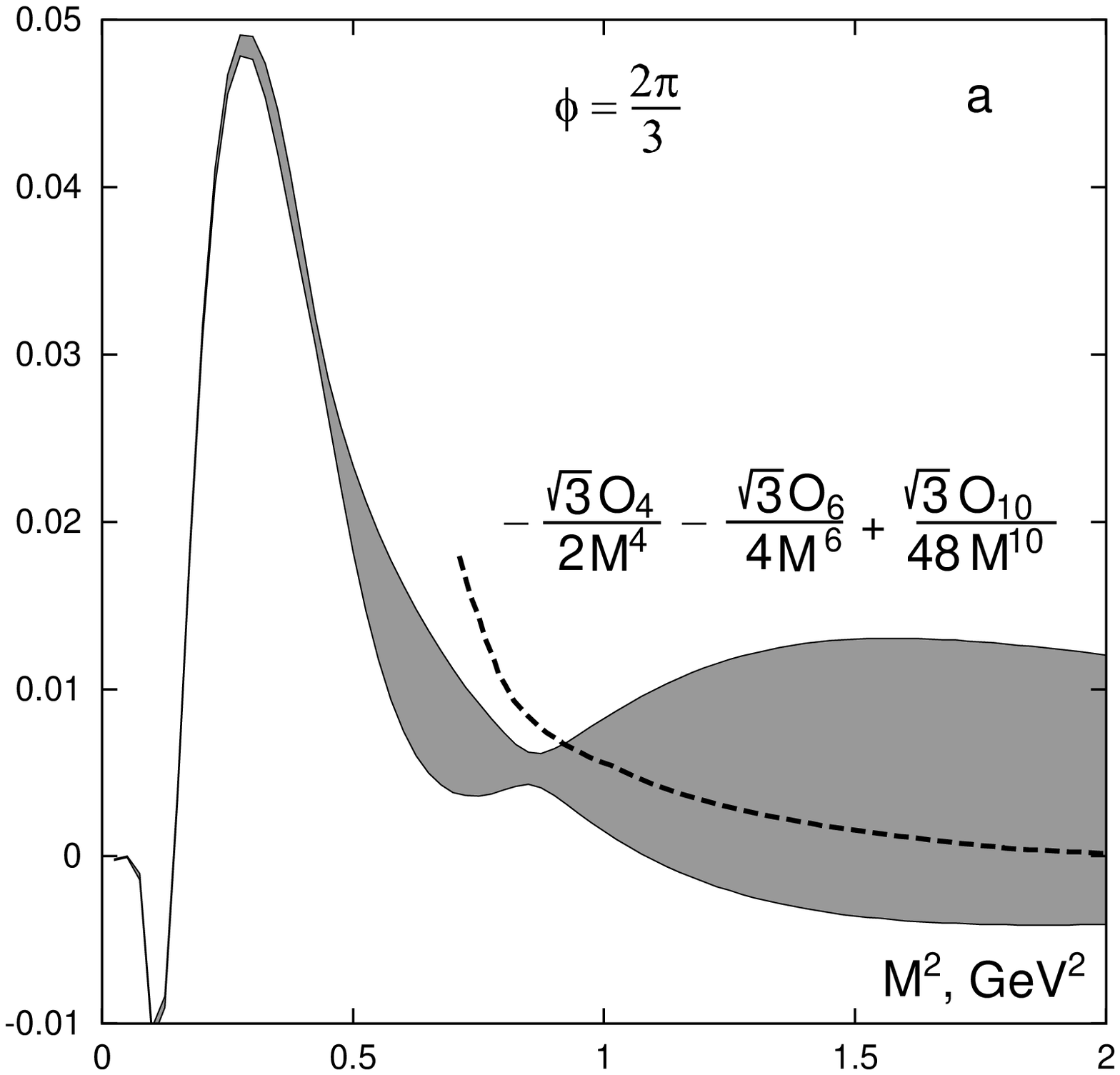, width=70mm} \hspace{5mm}
\epsfig{file=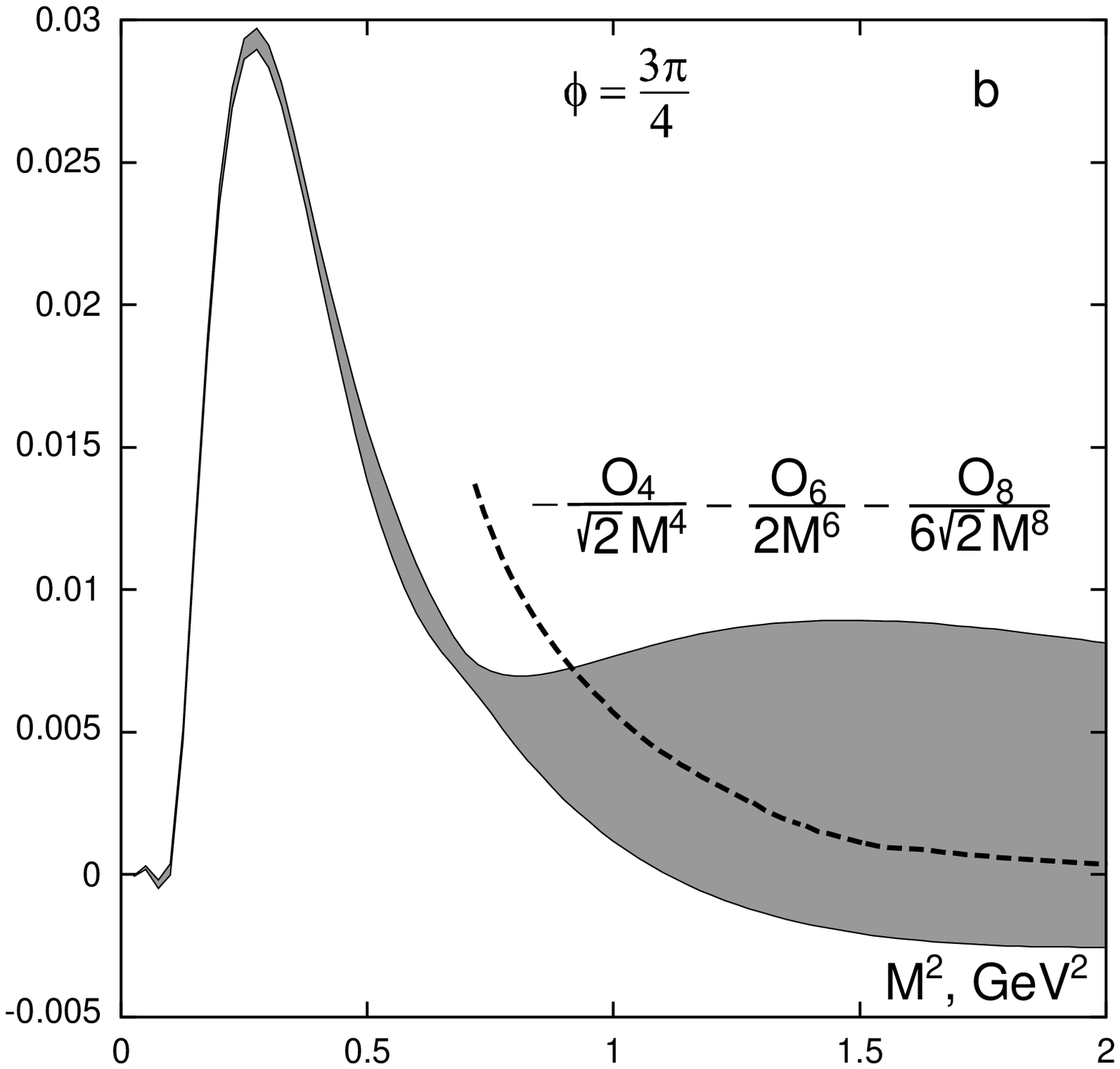, width=70mm} \caption{The same for
eq.(\ref{83}): a) $\phi = 2 \pi/3$, ~ b) $\phi = 3 \pi/4$. }
\end{figure}

When  comparing  the theoretical curves with experimental data it
must be taken in mind, that the value of $f_{\pi}$, which in the
figures was taken to be equal to experimental one $f_{\pi}=130.7
MeV$, in fact has a theoretical uncertainty of the order $(\Delta
f^2_{\pi}/f^2_{\pi})_{theor} \sim m^2_{\pi}/m^2_{\rho}$, where
$m_{\rho}$ is characteristic hadronic scale (say, $\rho$-meson
mass). This uncertainty is caused by chiral symmetry violation in
QCD. Particularly, the account of this uncertainty may lead to a
better agreement of theoretical curve with the data in Fig.6b. The
calculation of instanton contributions (Eq.(\ref{64})), shows,
that in all considered above cases they are less than $0.5\cdot
10^{-3}$ at $M^2 > 0.8~GeV^2$, i.e. are well below the errors. (In
some cases they improve the agreement with the data.) The best fit
of the data (the dashed curves at Fig.'s 6,7) was achieved at the
value
\be
O_6 = -4.4\cdot 10^{-3}~ GeV^6\label{84}\ee It follows from
(\ref{84}) after separating $\alpha_s$ correction
$[1+(89/48)\alpha_s/\pi)=1.33]$:
\be
\alpha_s\langle 0\mid \bar{q}q\mid 0\rangle^2 =1.5\cdot
10^{-4}~GeV^6\label{85}\ee The error may be estimated as 30\%. The
value (\ref{85}) in the limit of errors agrees with previous
estimation \cite{41}. The contribution of dimension 10 is
negligible in all cases at $M^2 \geq 1~ GeV^2$. It is worth
mentioning that the theory, i.e. the OPE agrees with the data at
$M^2 > 0.8 ~GeV^2$. The good agreement of the theoretical curves
with the data confirms the chosen value of $O_8$ (\ref{78}) and,
therefore, the use of factorization hypothesis. From (84), with
the use of $\alpha_s(1 GeV^2) = 0.55$ (see Fig.3)  the value of
quark condensate at 1 GeV can be found
\be
\langle 0 \vert qq \vert 0 \rangle_{1~ GeV} = -1.65 \cdot 10^{-2}~
GeV^3 = - (254~ MeV)^3 \label{85a} \ee and the convenient
parameter is
\be
a_{\bar{q}q} (1~GeV^2) \equiv - (2 \pi)^2 \langle 0 \vert \bar{q}
q \vert 0 \rangle_{1~GeV} = 0.65~ GeV^3 \label{86} \ee The
magnitude of quark condensate (86) is close to that which follows
from the Gell-Mann-Oakes-Renner relation (eq.(\ref{22})).

In the last years there were many attempts
\cite{43},\cite{b}-\cite{h} to determine quark condensates using
V-A spectral functions measured in $\tau$-decay. Unlike the
approach presented above, where the polarization operator
analytical properties were exploited in the whole complex
$q^2$-plane, what allowed one to separate the contribution of
operators of different dimensions, the authors of
\cite{43},\cite{b}-\cite{h} considered the finite energy sum rules
- FESR (or integrals over contours) with chosen weight functions.
In \cite{b,d} the $N_c \to \infty$ limit was used. In
\cite{c,d,e,h} an attempt was made to find higher dimension
condensates (up to 18 in [87], up to 16 in \cite{c,d} and up to 12
in \cite{e}). Determination of higher dimension condensates
requires fine tunning of the upper limit of integration in FESR.
If the upper limit of integration $s_0$ in FESR is below $2~GeV^2$
(e.g., such an upper limit, $s_0=1.47~GeV$ was chosen in
\cite{h}), then instanton-like corrections, not given by OPE are
of importance. (See Sec.4.2). The same remark refers to the case
of weight factors singular at $s=0$, like $s^{-l}$, $l > 0$
\cite{43}, when there is an enhancement of the contribution of low
$s$, where OPE breaks down. Taking in mind these remarks, we have
a satisfactory agreement of the values of condensate (\ref{84}),
presented above, with those found in \cite{43,d,h}.


\subsection{\it Determination of condensates from $V+A$ and $V$
structure functions of $\tau$-decay}

 Let us turn now to study  the $V+A$ correlator in the domain
of low $Q^2$, where the OPE terms play a much more essential role,
than in the determination of $R_\tau$. A general remark is in
order here. As was discussed in Ref.\cite{19} and stressed
recently by Shifman \cite{20}, the condensates cannot be defined
in rigorous way, because there is some arbitrariness in the
separation of their contributions from perturbative part. Usually
\cite{19,20} they are defined by introduction of some
normalization point $\mu^2$ with the magnitude of few $\Lambda^2_
{QCD}$. The integration over momenta in the domain below $\mu^2$
is addressed to condensates, above $\mu^2$ -- to perturbation
theory. In such formulation the condensates are $\mu$-dependent
$\left<O_D\right>=\left<O_D\right>_\mu$ and, strictly speaking,
they also depend on the way how the infrared cut-off $\mu^2$ is
introduced. The problem becomes more severe when the perturbative
expansion is performed up to higher order terms and the
calculation pretends on high precision. Mention, that this remark
does not refer to chirality violating condensates, because
perturbative terms do not contribute to chirality violating
structures in the limit of massless quarks. For this reason, in
principle, chirality violating condensates,
e.g.~$\bigl<0|\bar{q}q|0\bigr>$, can be determined with higher
precision, than chirality conserving ones.
 Here I use the definition of condensates,  which can be
called $n$-loop condensates.  As was formulated in Chapt.4, we
treat the renormalization group equation (\ref{54}) and the
equation for polarization operator (\ref{55}) in $n$-loop
approximation as exact ones; the expansion in inverse logarithms
is not performed. Specific values of condensates are referred to
such procedure. Of course, their numerical values depend on the
accounted number of loops; that is why the condensates, defined in
this way, are called $n$-loop condensates.

Consider the polarization operator
$\Pi=\Pi_{V+A}^{(1)}+\Pi_A^{(0)}$, defined in (\ref{50}) and its
imaginary part
\be
\omega(s)\,=\,v_1(s)\,+\,a_1(s)\,+\,a_0(s)\,=\,2\pi\, {\rm
Im}\,\Pi (s+i0) \label{87}\ee In parton model $\omega(s) \to 1$ at
$s\to \infty$. Any sum rule can be written in the following form:
\be
\int_0^{s_0} f(s) \, \omega_{\rm exp}(s)\,ds \,=\,i\pi \oint
f(s)\, \Pi_{\rm theor}(s) \,ds \label{88} \ee where $f(s)$ is some
analytical in the integration region function. In what follows we
use $\omega_{\rm exp}(s)$, obtained from $\tau$-decay invariant
mass spectra published in \cite{38} for $0<s<m_\tau^2$.  The
experimental error of the integral (\ref{88}) is computed as the
double integral with the covariance matrix
$\overline{\omega(s)\omega(s')}-\overline{\omega}(s)\overline{\omega}(s')$,
which also can be obtained from the data available in
Ref.\cite{38}. In the theoretical integral in (\ref{88}) the
contour goes from $s_0+i0$ to $s_0-i0$ counterclockwise around all
poles and cuts of theoretical correlator $\Pi(s)$. Because of
Cauchy theorem the unphysical cut must be inside the integration
contour.

The choice of the function $f(s)$ in Eq.(\ref{88}) is actually a
matter of taste. At first let us consider usual Borel
transformation:
\be
B_{\rm exp}(M^2) \,=\, \int_0^{m_\tau^2} e^{-s/M^2} \omega_{\rm
exp} (s) \, {ds\over M^2} \, = \, B_{\rm pt}(M^2) \,+\,2\pi^2
\sum_n {\bigl<O_{2n}\bigr>\over (n-1)! \, M^{2n}} \label{89} \ee
We separated out the purely perturbative contribution $B_{\rm
pt}$, which is computed numerically according to (\ref{88}) and
Eqs.(\ref{52})-(\ref{55}). Remind that Borel transformation
improves the convergence of OPE series because of the factors
$1/(n-1)!$ in front of operators and suppresses the contribution
of high-energy tail, where the experimental error is large. But it
does not suppress the unphysical perturbative cut, the main source
of the error in this approach, even increase it since
$e^{-s/M^2}>1$ for $s<0$. So the perturbative part $B_{\rm
pt}(M^2)$ can be reliably calculated only for $M^2 \ga 0.8-1\,
{\rm GeV}^2$ and higher; below this value the influence of the
unphysical cut is out of control.


\begin{figure}[tb]
\hspace{40mm} \epsfig{file=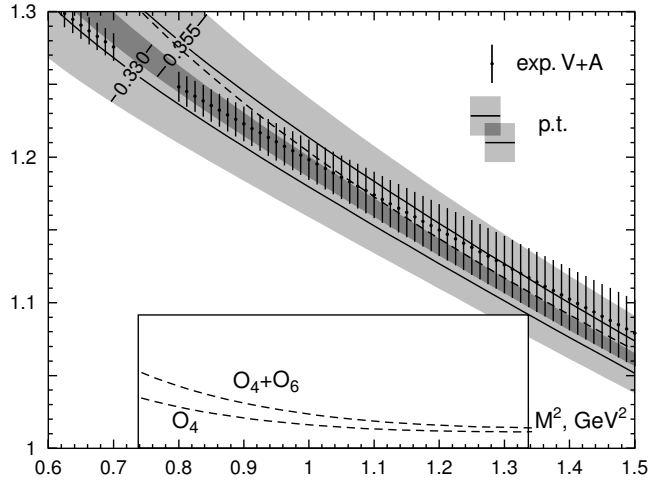, width=85mm} \caption{The
results of the Borel transformation of $V+A$ correlator for two
values $\alpha_s(m^2_{\tau}) = 0.355$ and $\alpha_s(m^2_{\tau}) =
0.330$. The widths of the bands correspond to PT errors, dots with
dashed errors -- experimental data. The dashed  curve is the sum
of the perturbative contribution at $\alpha_s(m^2_{\tau}) = 0.330$
and $O_4$, (Eq.'s (59),(\ref{91})) and $O_6$ (Eq.'s
(\ref{59}),(\ref{85})) condensate contributions.}
\end{figure}


Both $B_{\rm exp}$ and $B_{\rm pt}$ in 4-loop approximation for
$\alpha_s(m_\tau^2)=0.355$ and $0.330$ are shown in Fig.8. The
shaded areas display the theoretical error. They are taken equal
to the contribution of the last term in the perturbative Adler
function expansion $K_4 a^4$ (\ref{52}).

The contribution of the $O_8$ operator is of order
$O^{V-A}_8/N^2_c$ and negligible \cite{41}. (In fact, it depends
on the factorization procedure and uncertain for this reason). The
contributions of $D=4$ and $D=6$ operators are positive [see
(\ref{59})]. So, the theoretical perturbative curve must go below
the experimental points. The result shown in Fig.8 is in the
favour of the lower value of the QCD coupling constant
$\alpha_s(m^2_{\tau}) = 0.330$ (or, may be, $\alpha_s(m^2_{\tau})
= 0.340$). As is seen from Fig.8, the theoretical curve
(perturbative at $\alpha_s(m^2_{\tau}) =0.330$ plus OPE terms) is
in agreement with experiment at $M^2 \geq 0.9~ GeV^2$.

In order to separate the contribution of gluon condensate let us
perform the Borel transformation along the rays in the complex
$s$-plane in the same way, as it was done in Sec.5.1. The real
part of the Borel transform at $\phi = 5 \pi/6$ does not contain
$d=6$ operator.
\be
Re B_{exp} (M^2 e^{i 5 \pi/6}) = Re B_{pt}(M^2 e^{i 5 \pi/6}) +
\pi^2\frac{\langle O_4\rangle}{M^4} \label{90} \ee


\begin{figure}[tb]
\hspace{5cm} \epsfig{file=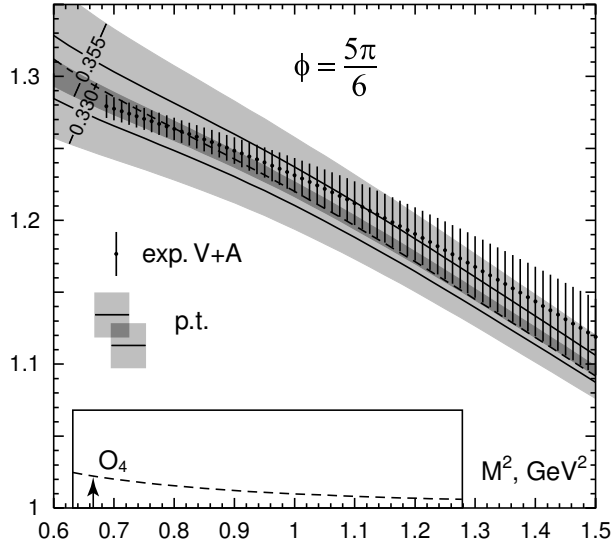, width=81mm} \caption{Real
part of the Borel transform (91) along the ray at the angle
$\phi=5\pi/6$ to the real axes. The dashed line corresponds to the
gluonic condensate given by the central value of (\ref{91}).}
\label{set1819fig}
\end{figure}

The results are shown in Fig.9. If we accept the lower value of
$\alpha_s(m^2_{\tau})$, we get the following restriction on the
value of gluon condensate:
\be
\left< {\alpha_s\over\pi} \,G_{\mu\nu}^a G_{\mu\nu}^a \right> \, =
\, 0.006\pm 0.012 \, {\rm GeV}^4 \; , \qquad
\alpha_s(m_\tau^2)=0.330 \quad {\rm and} \quad M^2 > 0.8 \, {\rm
GeV}^2 \label{91} \ee The theoretical and experimental errors are
added in quadratures  in Eq.(\ref{91}).

Turn now to analysis of the vector correlator (the vector spectral
function was published by ALEPH in  \cite{79}). In principle this
cannot give any new information in comparison with $V-A$ and $V+A$
cases. However the analysis of the vector current correlator is
important since it can also be performed with the experimental
data on $e^+e^-$ annihilation. The imaginary part of the
electromagnetic current correlator, measured there, is related to
the charged current correlator (\ref{45}) by the isotopic
symmetry. The statistical error in $e^+e^-$ experiments is less
than in $\tau$-decays because of significantly larger number of
events. So it would be interesting to perform similar analysis
with $e^+e^-$ data, which is a matter for separate research.

At first we consider usual Borel transformation for vector current
correlator, since it was originally applied in \cite{a2} for the
sum rule analysis.
 It is defined as (89) with the experimental
spectral function $\omega_{\rm exp}=2v_1$ instead of $v_1+a_1+a_0$
(the normalization is $v_1(s) \to 1/2$ at $s\to \infty$ in parton
model). Respectively, in the r.h.s. one should take the vector
operators $2O^V=O^{V+A}+O^{V-A}$. The  numerical results are shown
in  Fig.10. The perturbative theoretical curves are the same as in
Fig.8 with $V+A$ correlator. The dashed lines display the
contributions of the gluonic condensate given by Eq.(\ref{91}),
$2O_6^V=-3.5\times 10^{-3}\,{\rm GeV}^6$ and
$2O_8^V=O_8^{V-A}=-2.8\times 10^{-3} \,{\rm GeV}^8$ added to the
$\alpha_s(m^2_{\tau})=0.330$-perturbative curve. The contribution
of each condensate is shown in the box below. Notice, that for
such condensate values the total OPE contribution is small, since
positive $O_4$ compensate negative $O_6$  and $O_8$. The agreement
is observed for $M^2>0.8\,{\rm GeV}^2$.


\begin{figure}
\hspace{40mm} \epsfig{file=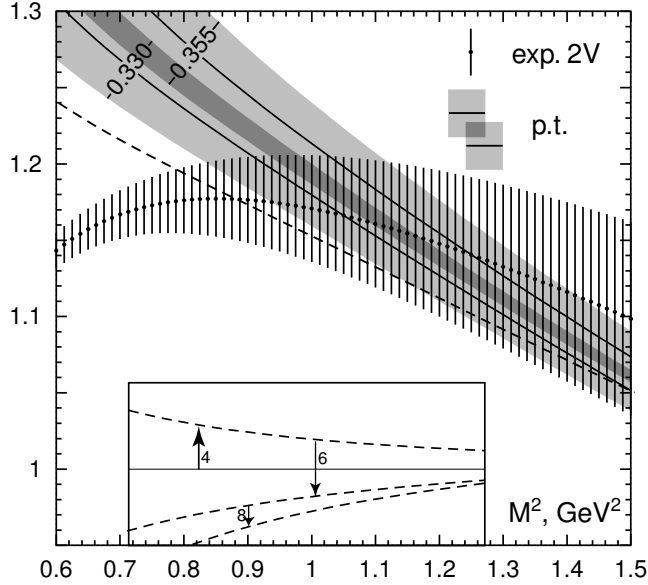, width=85mm} \caption{Borel
transformation for vector currents.} \label{borvfig}
\end{figure}

The Borel transformations along the rays in the complex plane
results in the same conclusion; at $M^2 > 0.8-0.9~ GeV^2$ the
agreement with experiment at 2\% level is achieved at
$\alpha_s(m^2_{\tau}) = 0.33-0.34$ and at the values of quark and
gluon condensates given by (84) and (\ref{91}). There is some
discrepancy in the vector spectral function obtained in
$\tau$-decay and in $e^+e^-$ annihilation (see \cite{44},
\cite{Hocker} and references herein): the $e^+e^-$ data are below
$\tau$-decay ones by 5-10\% in the interval $s = 0.6 - 0.8~
GeV^2$.  The substitution of $e^+e^-$ data instead of $\tau$-decay
data in the sum rule presented in Fig.10 does not spoil the
agreement of the theory with experiment in the limit of errors.

A few words about instanton contributions. They can be calculated
in the same way, as in the case of $V-A$ correlators. At the
chosen values of instanton gas parameters instanton contributions
are small, less than $0.5 \cdot10^{-3}$ at $M^2 > 0.8~ GeV^2$ and
do not spoil the agreement of the theory with experiment.

\bigskip

\section{Determination of quark condensate from QCD sum rules for
baryon masses}


Since in QCD with massless quarks the baryon masses arise due to
spontaneous violation of chiral symmetry and in a good
approximation, the proton mass (as well as $\Delta$-isobar) can be
expressed through quark condensate \cite{23}, the QCD sum rules
for baryon masses are a suitable tool for determination of quark
condensate assuming that baryon masses are known. The sum rules
can be derived by considering the polarization operator
\be
\Pi(p) = i~ \int~d^4 x e^{ipx} \langle 0 \vert T \{\eta(x),
\bar{\eta}(0)\} \vert 0 \rangle \label{92} \ee where $\eta(x)$ is
the quark current with baryon quantum numbers. In case of proton
the most suitable current is \cite{23,80}.
\be
\eta(x) = \varepsilon^{abc} (u^a C \gamma_{\mu} u^b) \gamma_5
\gamma_{\mu} d^c \label{93} \ee where $u^a$, $d^c$ -- are $u$ and
$d$ quark fields, $a,b,c$ are colour indeces, $C$ is the charge
conjugation matrix. After Borel transformation the sum rules for
proton mass have the form \cite{23,28,34}
$$ M^6 E_2 (s_0/M^2) L^{-4/9} \Biggl [1 + \Biggl (\frac{53}{12} +
\gamma_E) \frac{\alpha_s(M^2)}{\pi} \Biggr ] + \frac{1}{4} M^4 b
E_0 (s_0/M^2) L^{-4/9} + $$
\be
+\frac{4}{3} a^2_{\bar{q}q} \Biggl [1 + f(M^2)
\frac{\alpha_s(M^2)}{\pi} \Biggr ] - \frac{1}{3} a^2_{\bar{q}q}
\frac{m^2_0}{M^2} = \bar{\lambda}^2_p e^{-m^2/M^2} \label{94} \ee
\be
2 a_{\bar{q}q} M^4 E_1 (s_0/M^2) \Biggl [1 + \frac{3}{2}~
\frac{\alpha_s(M^2)}{\pi} \Biggr ] + \frac{272}{81}~
\frac{\alpha_s(M^2)}{\pi}~ \frac{a^3_{\bar{q}q}}{M^2} -
\frac{1}{12} a_{\bar{q}q}b = m \bar{\lambda}^2_p e^{-m^2/M^2}
\label{95}\ee Here $M$ is the Borel parameter, $m$ is the nucleon
mass, $a_{\bar{q}q}$ is given by (\ref{86}), $\gamma_E = 0.577$.
\be
b = (2 \pi)^2 \langle 0 \vert \frac{\alpha_s}{\pi}~ G^2 \vert 0
\rangle \label{96} \ee
 \be
 E_n(x) = \frac{1}{n!}~ \int\limits^{x}_{0}~ d z z^n e^{-z}
 \label{97}
 \ee
 \be
 L = \frac{\alpha_s(\mu^2)}{\alpha_s(M^2)},
 \label{98}
 \ee
 $L$ -- corresponds to anomalous dimensions, $s_0$ is the
 continuum threshold and $\mu^2$ is the normalization point,
 chosen as $\mu^2 = 1~GeV^2$. The constant $\bar{\lambda}_p$ is
 defined as $\bar{\lambda}^2_p = 2 (2 \pi)^4
 \lambda^2_{p}$
 \be
 \langle 0 \vert \eta \vert p \rangle = \lambda_p v_p,
 \label{99}
 \ee
where $v_p$ is the proton spinor. The $\alpha_s$ corrections to
proton sum rules were found in \cite{81}. The function $f(s)$ is
small, $\vert f \vert < 0.2$ at $0.9 < M^2< 1.5~ GeV^2$ and
$\alpha_s$ correction to the term proportional to $a^2_{\bar{q}q}$
can be neglected. The sum rules (\ref{94}), (\ref{95}) were
calculated at the following values of parameters: $\langle 0 \vert
(\alpha_s/\pi) G^2 \vert 0 \rangle = 0.005~ GeV^4$, ($b = 0.20~
GeV^4$), $s_0 = 2.5~ GeV^2$. The numerical value of quark
condensate was not fixed by the value given in (\ref{86}), but
considered as a free parameter. For the best fit of the sum rules
it was chosen to be $a_{\bar{q}q} = 0.60$ (cf.(\ref{86})). First,
the values of $\lambda^2_p$ was found from (\ref{94}),(\ref{95}),
where the experimental value of proton mass was substituted,
Fig.11, left scale. Then Eq.(\ref{95}) was divided by (\ref{94})
and the theoretical value of the proton mass was found, Fig.11,
right scale.


\begin{figure}
\hspace{45mm} \epsfig{file=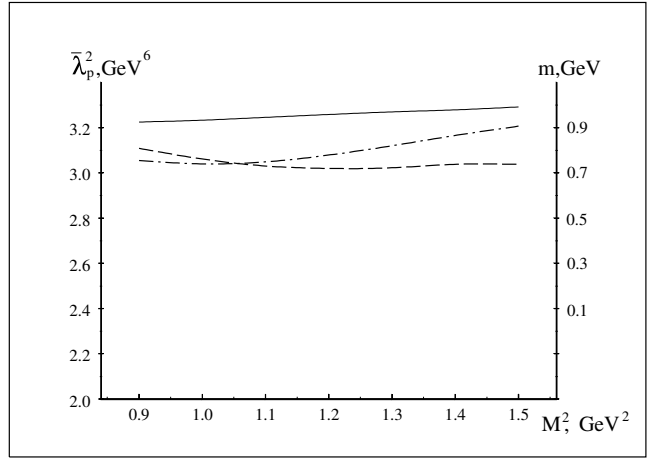, width=85mm} \caption{The
sum rules for proton mass Eq.s' (\ref{94}),(\ref{95}). The dashed
and dash-dotted curves give $\bar{\lambda}^2_p$, determined
correspondingly from (\ref{94}) and (\ref{95}), the experimental
value of $m$ was substituted (left scale). The solid line gives
$m$ as the ratio of (\ref{95}) to (\ref{94}). } \label{borvfig2}
\end{figure}

As is seen from Fig.11, $\bar{\lambda}^2_p$, determined  from
(\ref{94}),(\ref{95}) are almost independent on $M^2$ and coincide
with one another, as it should be. The proton mass value coincide
with the experimental one with a precision better than 3\%. The
conclusion is, that the value
\be
a_{\bar{q}q} = 0.60~ GeV^3 \label{100} \ee describes well the
proton mass sum rule. The main source of the error is the large
$\alpha_s$ correction (about 0.8) to the first term in (\ref{94}).
If we suppose that its uncertainty is 20\%, then the corresponding
error in $a_{\bar{q}q}$ is $\pm 0.1~ GeV^3$. Therefore, we get
from proton mass sum rules
\be
a_{\bar{q}q} = (0.60 \pm 0.10)~ GeV^3 \label{101}\ee


A remark about a possible role of instantons in  the sum rules for
proton mass. As was found in \cite{Dorokhov},\cite{Forkel} if the
quark current with proton quantum numbers is given by (\ref{94}),
then instantons do not change the sum rule (\ref{94}). Their
contribution to (96) is moderate in instanton gas model, if the
model parameters  are chosen as in (\ref{70})
\cite{Dorokhov,Forkel} and may shift the value of quark condensate
(102) by 10-20\%, i.e. in the limit of quoted error.

\bigskip

\section{Gluon condensate and determination of charmed quark mass
from charmonium spectrum}


\subsection{\it The method of moments. The results}


The existence of gluon condensate had been first demonstrated by
Shifman, Vainshtein and Zakharov \cite{1}. They considered the
polarization operator $\Pi_c(q^2)$ of the vector charmed current
\be
\Pi_c(q^2)(q_{\mu}q_{\nu} - \delta_{\mu \nu} q^2) = i~ \int~ d^4 x
e^{iqx} \langle 0 \vert T {J_{\mu}(x),~ J_{\nu}(0) } \vert 0
\rangle \label{102} \ee
\be
J_{\mu}(x) = \bar{c} \gamma_{\mu} c \label{103}\ee and calculated
the moments of $\Pi_c(q^2)$
\be
M_n(Q^2) = \frac{4 \pi^2}{n!} (-\frac{d}{d Q^2})^n \Pi_c(Q^2),
\label{104} \ee $(Q^2 = -q^2)$ at $Q^2 = 0$. The OPE for
$\Pi(Q^2)$ was used and only one term in OPE series was accounted
-- the gluonic condensate. In perturbative part of $\Pi(Q^2)$ only
the first order term in $\alpha_s$ was accounted and a small value
of $\alpha_s$ was chosen, $\alpha_s(m_c) \approx 0.2$. The moments
were saturated by contribution of charmonium states and in this
way the value of gluon condensate (\ref{27}) was found. The SVZ
approach \cite{1} was criticized in \cite{82}, where it was shown
that the higher order terms of OPE, namely, the contributions of
$G^3$ and $G^4$ operators are of importance at $Q^2 = 0$.
Reinders, Rubinstein and Yazaki \cite{83} demonstrated, however,
that SVZ results may be restored, if one considers not small
values $Q^2 > 0$ instead of $Q^2=0$.
 Later there were many attempts to determine the gluon
  condensate by considering various processes
within various approaches. In some of them the value (\ref{27})
(or ones, by a factor of $1.5$ higher) was confirmed
\cite{a2,84,85}, in others it was claimed, that the actual value
of the gluon condensate is by a factor 2--5 higher than (\ref{27})
\cite{86}.

From today's point of view the calculations performed in \cite{1}
have a serious drawback. Only the first order (NLO) perturbative
correction was accounted in \cite{1} and it was taken rather low
value of $\alpha_s$, later not confirmed by the experimental data.
The contribution of the next, dimension 6, operator $G^3$ was
neglected, so the convergence of the operator product expansion
was not tested.

There are recent publications \cite{87} where the charmonium as
well as bottomonium sum rules were analyzed at $Q^2=0$ with the
account of  $\alpha_s^2$ perturbative corrections in order to
extract the charm and bottom quark masses in various schemes. The
condensate is usually taken
 to be 0 or some another fixed value. However, the charm mass and the
  condensate
values are entangled in the sum rules. This can be easily
understood for large $Q^2$, where the mass and condensate
corrections to the polarization operator behave as some series in
negative powers of $Q^2$, and one may eliminate the condensate
contribution to a great extent by slightly changing the quark
mass.  Vice versa, different condensate values may vary the charm
quark mass within few per cents. (See Fig.12 below.)

 Therefore, in order to perform reliable calculation of gluon
 condensate by studying the moments of charmed current
 polarization operator it is necessary to account $\alpha^2_s$
 perturbative corrections to the moments, $\alpha_s$ corrections
 to gluon condensate contribution, $\langle G^3\rangle$ term in OPE
 and to find the region in $(n,Q^2)$ space, where all these
 corrections are small. This program was realized in Ref.\cite{88}.
 The basic points of this consideration are presented below.

The dispersion representation for $\Pi(q^2)$ has the form
\be
R(s)\,=\,4\pi \, {\rm Im} \, \Pi_c(s+i0) \; , \qquad
\Pi_c(q^2)\,=\,{q^2\over 4\pi^2}\int_{4m^2_c}^\infty
\,{R(s)\,ds\over s(s-q^2)} \; , \label{105}\ee where $R(\infty) =
1$ in partonic model. In approximation of infinitely narrow widths
of resonances $R(s)$ can be written as a sum of contributions from
resonances and continuum
\be
\label{rexp} R(s)\,=\,{3 \, \pi \over Q_c^2 \, \alpha_{\rm
em}^2\!(s)}\, \sum_\psi m_\psi \Gamma_{\psi \to
ee}\,\delta(s-m_\psi^2) \,+\,\theta(s-s_0) \label{106}\ee where
$Q_c = 2/3$ is the charge of charmed quarks, $s_0$ - is the
continuum threshold (in what follows $\sqrt{s_0} = 4.6~ GeV$), ~~
$\alpha(s)$ - is the running electromagnetic constant,~
$\alpha(m^2_{J/\psi}) = 1/133.6$.  The polarization operator
moments are expressed through $R$ as:
\be
M_n(Q^2)=\int_{4m^2_c}^\infty {R(s)\, ds\over (s+Q^2)^{n+1}}
\label{107}\ee According to (\ref{107}) the experimental values of
moments are determined by the equality
\be
\label{momexp} M_n(Q^2)\,=\,{27\,\pi\over 4\, \alpha_{\rm
em}^2}\sum_{\psi=1}^6 {m_\psi\Gamma_{\psi\to ee}\over
(m_\psi^2+Q^2)^{n+1}} \,+\,{1\over n (s_0+Q^2)^n} \label{108}\ee
In the sum in (\ref{108}) the following resonances were accounted:
$J/\psi(1S)$, $\psi(2S)$, $\psi(3770), \psi(4040)$,
$\psi(4160),\psi(4415)$, their $\Gamma_{\psi\to ee}$ widths were
taken from PDG data \cite{10}. It is reasonable to consider the
ratios of moments $M_{n1}(Q^2)/M_{n2}(Q^2)$ from which the
uncertainty due to error in $\Gamma_{J/\psi \to ee}$ markedly
falls out. Theoretical value for $\Pi(q^2)$ is represented as a
sum of perturbative and nonperturbative contributions. It is
convenient to express the perturbative contribution through
$R(s)$, making use of (\ref{105}),(\ref{107}):
\be
R(s)\,=\,\sum_{n\ge 0} R^{(n)}(s,\mu^2)\, a^n(\mu^2)
\label{109}\ee where $a(\mu^2) = \alpha_s(\mu^2)/\pi$. Nowadays,
three terms of expansion in (\ref{109}) are known: $R^{(0)}$
\cite{89} $R^{(1)}$ \cite{90},  $R^{(2)}$ \cite{91}.  They are
represented as functions of quark velocity $v = \sqrt{1 -
4m^2_c/s}$, ~ where $m_c$ -- is the pole mass of quark. Since they
are cumbersome, I will not present them here (see \cite{88} for
details).

Nonperturbative contributions into polarization operator have the
form (restricted by d=6 operators):
$$\Pi_{nonpert}(Q^2) = \frac{1}{(4m^2_c)^2} \langle 0\mid
\frac{\alpha_s}{\pi} G^2 \mid 0 \rangle [~f^{(0)}(z) +af^{(1)}
(z)~] + $$
\be
+\frac{1}{(4m^2_c)^3} g^3 f^{abc} \langle 0 \mid G^a_{\mu\nu}
G^b_{\nu\lambda} G^c_{\lambda \mu} \mid 0 \rangle F(z),~~
z=-\frac{Q^2}{4m^2_c}\label{110}\ee

 Functions $f^{(0)}(z)$, ~ $f^{(1)}(z)$ and $F(z)$ were
calculated in \cite{1}, \cite{92}, \cite{93}, respectively. The
use of the quark pole mass is, however, inacceptable. The matter
is that in this case the PT corrections to moments are very large
in the region of interest and perturbative series  seems to
diverge.

So, it is reasonable to use $\overline{MS}$ mass
$\overline{m}(\mu^2)$, taken at the point $\mu^2 =
\overline{m}^2$. The calculations, performed in ref.100 show, that
in the region near the diagonal in $(Q^2,n)$ plane, $Q^2/4m^2 =
n/5 - 1$ all mentioned above corrections are small. For example,

\be
n=10 \; , \; Q^2=4{\bar m}^2_c:  \qquad {{\bar M}^{(1)}\over {\bar
M}^{(0)}}=0.045 \; , \qquad {{\bar M}^{(2)}\over {\bar
M}^{(0)}}=1.136 \; , \qquad {{\bar M}^{(G,1)}\over {\bar
M}^{(G,0)}}=-1.673 \label{111} \ee (here $M^{(k)}$ mean the
coefficients at the contributions of terms $\sim a^k$ to the
moments, $M^{(G,k)}$ - are the similar coefficients for gluonic
condensate contribution).

At $a \sim 0.1$ and at the ratios of moments given by (\ref{111})
there is a good reason to believe that the PT series well
converges. Such a good convergence holds (at $n > 5$) only in the
case of large enough $Q^2$, at $Q^2 = 0$  one does not succeed in
finding such $n$, that perturbative corrections to the moments,
$\alpha_s$ corrections to gluonic condensates and the term $\sim
\langle G^3 \rangle$ contribution would be simultaneously small.


\begin{figure}[tb]
\hspace{40mm} \epsfig{file=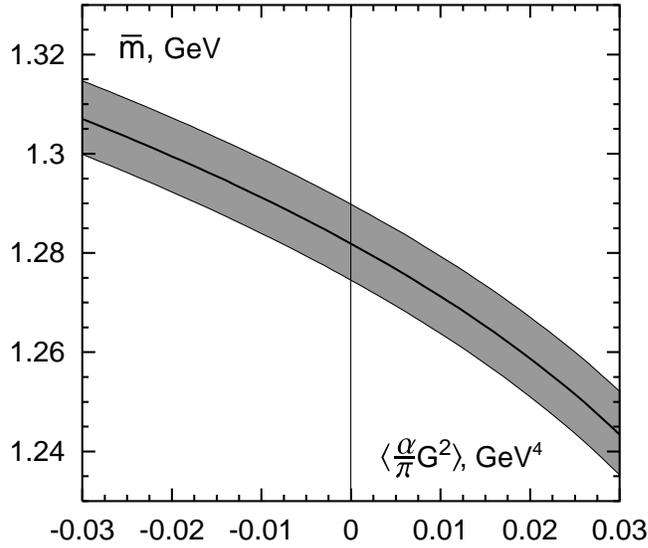, width=85mm} \caption{The
dependence of $\overline{m}(\overline{m})$ on $\langle 0 \vert
\alpha_s/\pi)G^2 \vert 0 \rangle$ obtained at $n = 10$,~ $Q^2 =
0.98 \cdot 4m^2$ and $\alpha_s(Q^2 + \overline{m}^2)$. }
\end{figure}

\begin{figure}[tb]
\hspace{10mm} \epsfig{file=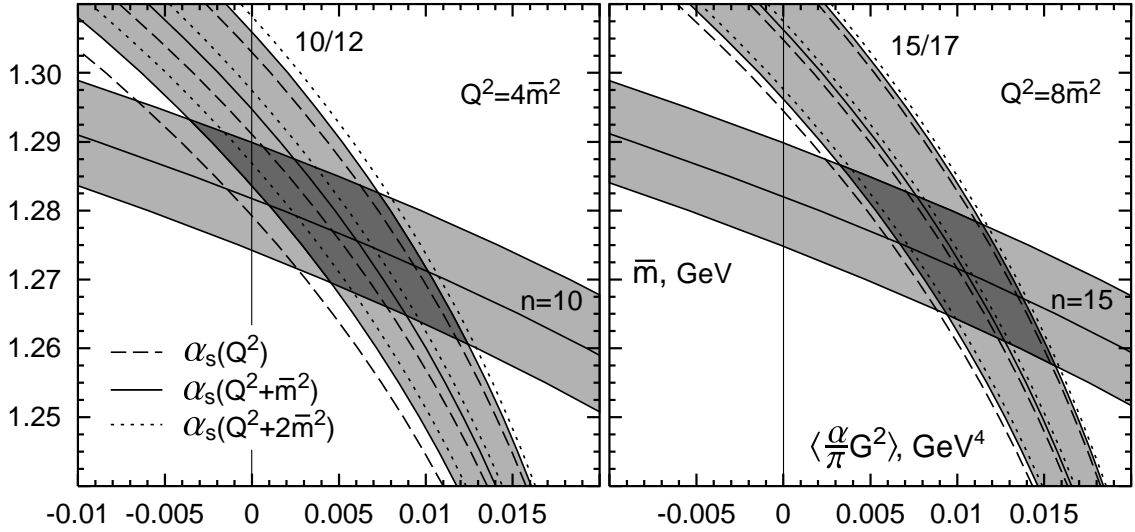, width=150mm} \caption{The
dependence of $\overline{m}(\overline{m})$ on $\langle 0 \vert
(\alpha_s/\pi) G^2 \vert 0 \rangle$ obtained from the moments
(horizontal bands) and their ratios (vertical bands) at different
$\alpha_s$. The left-hand figure: $Q^2 = 4 \overline{m}^2$, ~ $n =
10$, ~ $M_{10}/M_{12}$; the right-hand figure -- $Q^2 = 8
\overline{m}^2$, ~ $n = 15$, ~ $M_{15}/M_{17}$. }
\end{figure}

It is also necessary to choose the scale - normalization point
$\mu^2$ where $\alpha_s(\mu^2)$ is taken. In (\ref{109})~ $R(s)$
is a physical value and cannot depend on $\mu^2$. Since, however,
we take into account in (\ref{109}) only three terms, at
unsuitable choice of $\mu^2$ such $\mu^2$ dependence may arise due
to neglected terms. At large $Q^2$  the natural choice is $\mu^2 =
Q^2$. It can be thought that at $Q^2 = 0$ the reasonable scale is
$\mu^2 = \overline{m}^2$, though some numerical factor is not
excluded in this equality. That is why it is reasonable to take
interpolation form
\be
\mu^2 = Q^2+\overline{m}^2,\label{112}\ee but to check the
dependence of final results on a possible factor at
$\overline{m}^2$. Equalling theoretical value of some moment at
fixed  $Q^2$ (in the region where $M^{(1)}_n$ and $M^{(2)}_n$ are
small) to its experimental value one can find the dependence of
$\overline{m}$ on $\langle(\alpha_s/\pi)G^2 \rangle$ (neglecting
the terms $\sim \langle G^3 \rangle$). Such a dependence for $n =
10$ and $Q^2/4 m^2 = 0.98$ is presented in Fig.12.

To fix both $\overline{m}$ and $\langle(\alpha_s/\pi) G^2 \rangle$
one should, except for moments, take their ratios. Fig.13 shows
the value of $\overline{m}$ obtained from the moment $M_{10}$ and
the ratio $M_{10}/M_{12}$ at $Q^2 = 4 m^2$ and from the moment
$M_{15}$ and the ratio $M_{15}/M_{17}$ at $Q^2 = 8 m^2$. The best
values of masses of charmed quark and gluonic condensate are
obtained from fig.13:
\be
{\bar m}({\bar m}^2)\,=\,1.275\pm 0.015 \, {\rm GeV} \; , \qquad
\left< {\alpha_s\over \pi} G^2\right>\,=\,0.009\pm 0.007 \, {\rm
GeV}^4 \label{113}\ee  The calculation shows, that the influence
of continuum -- the last term in eq.(\ref{106}) is completely
negligible. Up to now the corrections $\sim \langle G^3 \rangle$
were not taken into account. It appears that in the region of $n$
and $Q^2$ used to find $\overline{m}$ and gluonic condensate they
are comparatively small and, practically, not changing
$\overline{m}$, increase $\langle (\alpha_s/\pi)G^2\rangle $ by
$10-20\%$ if the term $\sim \langle G^3 \rangle$ is estimated
according to (\ref{29}) at $\rho_c = 0.5 fm$.

It should be noted that improvement of the accuracy of
$\Gamma_{J/\psi \to ee}$ would make it possible to precise the
value of gluonic condensate: the widths of horizontal bands in
fig.13 are determined mainly just by this error. In particular,
this, perhaps, would allow one to exclude the zero value of
gluonic condensate, that would be extremely important.
Unfortunately, eq.(114) does not allow one to do it for sure.
Diminution of theoretical errors which determine the width of
vertical bands seems to be less real.

In order to check the results (114) for gluon condensate the
pseudoscalar and axial-vector channels in charmonia were
considered. The same method of moments was used and the regions in
the space $(n,Q^2)$ were found, where higher order perturbative
and OPE terms are small. In the pseudoscalar case it was obtained
\cite{94} that, if for $\overline{m}$ the value (114) is accepted
and the contribution of $\langle 0 \vert G^4 \vert 0 \rangle$
condensate may be neglected, then there follows the upper limit
for gluon condensate
\be
\langle 0 \vert \frac{\alpha_s}{\pi}~ G^2 \vert 0 \rangle < 0.008~
GeV^4 \label{114} \ee The contribution of $D=6$ condensate
$\langle 0 \vert G^3 \vert 0 \rangle$ is shown to be small. If
$\langle G^4 \rangle$ condensate is accounted and its value is
estimated by factorization hypothesis, then the upper limit for
gluon condensate increases to
\be
\langle 0 \vert \frac{\alpha_s}{\pi}~ G^2 \vert 0 \rangle < 0.015~
GeV^4 \label{115} \ee

In \cite{95} the case of the axial-vector channel in charmonia was
investigated and very strong limitations on gluon condensate were
found:
\be
\langle 0 \vert \frac{\alpha_s}{\pi}~ G^2 \vert 0 \rangle =
0.005^{+0.001}_{-0.004}~ GeV^4 \label{116} \ee Unfortunately,
(\ref{116}) does not allow one to exclude the zero value for gluon
condensate. It should be mentioned, that the allowed region in
$(n,Q^2)$ space, where all corrections are small, is very narrow
in this case, what does allow us in \cite{95} to check the result
(\ref{116}) by studying some other regions in $(n,Q^2)$, as it was
done in the two previous cases -- vector and pseudoscalar.

Let us now turn the problem around and try to predict the width
$\Gamma_{J/\psi\to ee}$ theoretically. In order to avoid the wrong
circle argumentation we do not use the condensate value just
obtained, but take the limitation $\left< {\alpha_s\over
\pi}G^2\right>=0.006\pm 0.012 \, {\rm GeV}^4$ found  from
$\tau$-decay data. Then, the mass limits ${\bar m}=1.28
-1.33\,{\rm GeV}$ can be found from the moment ratios exhibited
above, which do not depend on $\Gamma_{J/\psi\to ee}$ if the
contributions of higher resonances is approximated by continuum
(the accuracy of such approximation is about $3\%$). The
substitution of these values of ${\bar m}$ into the moments gives
\be
 \Gamma^{\rm theor}_{J/\psi\to ee}\,=\,4.9 \pm 0.8 \, {\rm
keV}\label{117} \ee in comparison with experimental value
$\Gamma_{J/\psi\to ee}=5.26\pm 0.37\,{\rm keV}$. Such a good
coincidence of the theoretical prediction with experimental data
is a very impressive demonstration of the QCD sum rules
effectiveness. It must be stressed, that while obtaining
(\ref{117}) no additional input were used besides the condensate
restriction taken from Eq.(\ref{91}) and the value of
$\alpha_s(m_\tau^2)$.

\subsection{\it The attempts to sum up the Coulomb-like corrections.
Recent publications}


Sometimes when considering  the heavy quarkonia sum rules the
Coulomb-like corrections are summed up   \cite{96}-\cite{100}. The
basic argumentation for such summation is that at $Q^2=0$ and high
$n$ only small quark velocities $v\la 1/\sqrt{n}$ are essential
and the problem becomes nonrelativistic. So it is possible to
perform the summation with the help of well known formulae of
nonrelativistic quantum mechanics for $|\psi(0)|^2$ in case of
Coulomb interaction (see \cite{101}).

This method was not used here for the following reasons:

1. The basic idea of our approach is to calculate the moments of
the polarization operator in QCD by applying the perturbation
theory and OPE (l.h.s.~of the sum rules) and to compare it with
the r.h.s.~of the sum rules, represented by the contribution of
charmonium states (mainly by $J/\psi$ in vector channel).
Therefore it is assumed, that the theoretical side of the sum rule
is dual to experimental one, i.e. the same domains of coordinate
and momentum spaces are of importance at both sides. But the
charmonium states (particularly, $J/\psi$) are by no means the
Coulomb systems. A particular argument in favor of this statement
is the ratio $\Gamma_{J/\psi\to ee}/\Gamma_{\psi'\to ee}=2.4$. If
charmonia were nonrelativistic Coulomb system, $\Gamma_{\psi\to
ee}$ would be proportional to $|\psi(0)|^2\sim 1/(n_r+1)^3$, and
since $\psi'$ is the first radial excitation with $n_r=1$, this
ratio would be equal to 8 (see also \cite{101}).

2. The heavy quark-antiquark Coulomb interaction at large
distances $r>r_{\rm conf}\sim 1 \,{\rm GeV}^{-1}$ is screened by
gluon and light quark-antiquark clouds, resulting in string
formation. Therefore the summation of Coulombic series makes sense
only when the Coulomb radius $r_{\rm Coul}$ is below $r_{\rm
conf}$. (It must be taken in mind, that higher order terms in
Coulombic series represent the contributions of large distances,
$r\gg r_{\rm Coul}$.) For charmonia we have \be r_{\rm
Coul}\,\approx\, {2\over m_c C_F \alpha_s}\, \approx  \, 4 \,{\rm
GeV}^{-1} \label{118}\ee It is clear, that the necessary condition
$R_{\rm Coul} < R_{\rm conf}$ is badly violated for charmonia.
This means that the summation of the Coulomb series in case of
charmonium would be a wrong step.

3. The analysis is performed at $Q^2/4{\bar m}^2\ge 1$. At large
$Q^2$ the Coulomb corrections are suppressed in comparison with
$Q^2=0$. It is easy to estimate the characteristic values of the
quark velocities. At large $n$ they are
$v\approx\sqrt{(1+Q^2/4m^2)/n}$. In the region $(n,Q^2)$ the
exploited above quark velocity $v\sim 1/\sqrt{5}\approx 0.45$ is
not small
 and not in the nonrelativistic domain, where the Coulomb corrections
 are large and legitimate.

 Nevertheless let us look on the expression of $R_c$, obtained after
 summation of the Coulomb corrections in the
 nonrelativistic theory \cite{102}.
 It reads (to go from QED to QCD one has to replace $\alpha\to C_F\alpha_s$,
 $C_F=4/3$):
 \be
 R_{c, \,{\rm Coul}}\,=\,{3\over 2} \, {\pi C_F \alpha_s\over 1 - e^{-x}}  \,=\,
 {3\over 2} \, v \left( \, 1\,+ \,{x\over 2}\,+ \,{x^2\over 12} \,-\,{x^4
 \over 720}\,+\,\ldots\, \right)
 \label{119}
 \ee
 where $x=\pi C_F\alpha_s/v$. At $v= 0.45$
 and $\alpha_s\approx 0.26$ the first 3 terms in the expansion
 (\ref{119}), accounted in our calculations, reproduce the exact value of
  $R_{c,\, {\rm Coul}}$ with the accuracy
 $1.6\%$. Such deviation leads to the error of the mass ${\bar m}$ of order
  $(1-2)\times 10^{-3}\,{\rm GeV}$,
 which is completely negligible. In order to avoid misunderstanding,
 it must be mentioned, that the value of
 $R_{c\, {\rm Coul}}$, computed by summing the Coulomb corrections in
 nonrelativistic theory has not too
 much in common with the real physical situation. Numerically, at chosen
  values of the parameters,
 $R_{c\, {\rm Coul}}\approx 1.8$, while the real value (both experimental
 and
 in the perturbative QCD)
 is about $1.1$. The goal of the arguments, presented above, was
 to demonstrate, that even in the case of
 Coulombic system  our approach would have a good accuracy of calculation.

At $v=0.45$ the momentum transfer from quark to antiquark
 is $\Delta p \sim 1\,{\rm GeV}$. (This is a typical domain for
QCD sum rule validity.) In coordinate space it corresponds to
$\Delta r_{q {\bar q}} \sim 1\, {\rm GeV}^{-1}$. Comparison with
potential models \cite{102} demonstrates, that in this region the
effective potential strongly differs from Coulombic one.

4. Large compensation of various terms in the expression for the
moments in $\overline{\rm MS}$ scheme  is not achieved, if only
the Coulomb terms are taken into account. This means, that the
terms of non-Coulombic origin are more important here, than
Coulombic ones.

For all these reasons   the summation of nonrelativistic Coulomb
corrections is inadequate in the problem in view: it will not
improve the accuracy of calculations, but would be misleading.

In the recent publication \cite{103} it is claimed, that gluon
condensate is much larger than the presented above values, it was
found $\langle 0 \vert (\alpha_s/\pi)~ G^2 \vert 0 \rangle = 0.062
\pm 0.019~ GeV^4$. The author of \cite{103} considered the model,
where hadronic  spectrum is represented by infinite number of
vector mesons. The polarization operator, calculated in this model
was equalled to those in QCD, given by perturbative and OPE terms.
The value of gluon condensate was found from this equality. The
zero width approximation was used for vector mesons. It is clear,
however, that the account of non-zero widths results in the terms
of the same type, proportional to $1/Q^4$, as the contribution of
gluon condensate. The sign of these terms is such, that they lead
to diminishing of gluon condensate. Namely, after accounting for
$\rho$-meson width, the value of gluon condensate decreases by a
factor of 2. For this reason the results of \cite{103} are not
reliable.

\bigskip

 \section{Valence quark distributions in nucleon at low $Q^2$
and the condensates }

\hspace{7mm} Quark and gluon distributions in hadrons are not
fully understood in QCD. QCD predicts the evolution of these
distributions with $Q^2$ in accord with the
Dokshitzer-Gribov-Lipatov-Altarelli-Parisi (DGLAP)
\cite{1a}-\cite{3a} equations, but not the initial values from
which this evolution starts. The standard way of determination of
quark and gluon distributions in nucleon is the following
\cite{4a}-\cite{8a} (for the recent review see \cite{9a}). At some
$Q^2 = Q^2_0$ (usually, at low or intermediate $Q^2 \sim 2-5
GeV^2$) the form of quark (valence and sea) and gluon
distributions is assumed and characterized by the number of free
parameters. Then, by using DGLAP equations, quark and gluon
distributions are calculated at all $Q^2$ and $x$ and compared
with the whole set of the data on deep inelastic lepton-nucleon
scattering (sometimes also with prompt photon production, jets at
high $p_{\bot}$ etc). The best fit for the parameters is found
and, therefore, quark and gluon distributions are determined at
all $Q^2$, including their initial values $q(Q^2_0, x)$,
~$g(Q^2_0, x)$. Evidently, such an approach is not completely
satisfactory from theoretical point of view - it would be
desirable to determine the initial distribution directly from QCD.
In QCD calculation valence quark distributions in nucleon
essentially depend on vacuum condensate, particularly, on gluon
condensate. Therefore, the comparison of valence quark
distributions calculated in QCD with those , found by the fit to
the data, allows one to check the values of condensates obtained
by consideration of quite different physical phenomena. For all
these reasons it is desirable to find quark and gluon distribution
in nucleon at low $Q^2 \sim 2-5 GeV^2$ basing directly on QCD.

The method of calculation of valence quark distributions at low
$Q^2(Q^2=2-5~GeV^2)$  was suggested in \cite{11a} and developed in
\cite{12a}-\cite{14a}. Recently, the method  had been improved and
valence quark distributions in pion \cite{15a} and transversally
and longitudinally polarized $\rho$-meson \cite{10a} had been
calculated, what was impossible in the initial version of the
method. The idea of the approach (in the improved version) is to
consider the imaginary part (in $s$-channel) of a four-point
correlator $\Pi(p_1, p_2, q, q^{\prime})$ corresponding to the
non-forward scattering of two quark currents, one of which has the
quantum numbers of hadron of interest (in our case -- of proton)
and the other is electromagnetic (or weak). It is supposed that
virtualities of the photon  $q^2, q^{\prime 2}$ and hadron
currents $p^2_1, p^2_2$ are large and negative $\vert q^2 \vert =
\vert q^{\prime 2} \vert \gg \vert p^2_1 \vert,~ \vert p^2_2 \vert
\gg R^{-2}_c$, where $R_c$ is the confinement radius. It was shown
in \cite{11a} that in this case the imaginary part in $s$-channel
$[s = (p_1 + q)^2]$ of $\Pi(p_1, p_2; q, q^{\prime})$ is dominated
by a small distance contribution at intermediate $x$. (The
standard notation is used: $x$ is the Bjorken scaling variable, $x
= -q^2/2 \nu$~, $\nu = p_1 q)$. The proof of this statement is
given in ref.\cite{12a}. So, in the mentioned above domain of
$q^2, q^{\prime 2}$,~ $p^2_1, p^2_2$ and intermediate $x$~ $Im
\Pi(p,p_2; q, q^{\prime})$ can be calculated using the
perturbation theory and the operator product expansion in both
sets of variables $q^2 = q^{\prime 2}$ and $p^2_1, p^2_2$. Only
the lowest twist terms, corresponding to condition $\mid
p^2_1/q^2\mid \ll 1$, $\mid p^2_2/q^2\mid \ll 1$ are considered.

The approach is inapplicable at small $x$ and $x$ close to 1. This
can be easily understood for physical reasons.  In deep inelastic
scattering at large $\vert q^2 \vert$ the main interaction region
in space-time is the light-cone domain and longitudinal distances
along the light-cone are proportional to $1/x$ and become large at
small $x$ \cite{16a,17a}. For OPE validity it is necessary for
these longitudinal distances along light-cone to be also small,
that is not the case at small $x$. At $1 - x \ll 1$ another
condition of applicability of the method is violated. The total
energy square $s = Q^2(1/x - 1) + p^2_1$~ $Q^2 = -q^2$ is not
large at $1 - x \ll 1$. Numerically, the typical values to be used
below are  $Q^2 \sim 5~ GeV^2$, ~ $p^2_1 \sim - 1~ GeV^2$. Then,
even at $x \approx 0.7$,~~ $s \approx 1~ GeV^2$, i.e., at such $x$
we are in the resonance, but not in the scaling region. So, one
may expect beforehand, that our method could work only up to $x
\approx 0.7$. The inapplicability of the method at small and large
$x$ manifests itself in the blow-up of higher order terms of OPE.
More precise limits on the applicability domain in $x$ will be
found from the magnitude of these terms.

The further procedure is common for QCD sum rules. On one hand the
four-point correlator $\Pi(p_1, p_2; q, q^{\prime})$ is calculated
by perturbation theory and OPE.On the other hand, the double
dispersion representation in $p^2_1, p^2_2$ in terms of physical
states contributions is written for the same correlator and the
contribution of the lowest state is extracted using the Borel
transformation. By equalling these two expression the desired
quark distribution is found. Valence quark distributions in proton
according to this method were calculated in \cite{A}. The basic
results of \cite{A} are presented below.

Consider the 4-current correlator which corresponds to the virtual
photon scattering on the quark current with quantum number of
proton:

$$ T^{\mu \nu} (p_1, p_2, q, q^{\prime}) = - i \int d^4 x d^4 y
d^4 z \cdot e^{i(p_1 x+ q y - p_2 z)}\cdot $$
\be
\cdot \langle 0 \vert T \{ \eta (x),~ j^{u, d}_{\mu} (y), j^{u,
d}_{\nu} (0), ~ \bar{\eta} (z) \} \vert 0 \rangle,\label{120} \ee
where $\eta(x)$ is the three-quark current (\ref{93}). Choose the
currents in the form $j^u_{\mu} = \bar{u} \gamma_{\mu}u$,~
$j^d_{\mu} = \bar{d} \gamma_{\mu} d$, i.e. as an electromagnetic
current which interacts only with  $u(d)$ quark (with unit
charges). Such a choice allows us to get sum rules separately for
distribution functions of $u$ and $d$ quarks. Let us take the
hadronic currents momenta to be nonequal, perform the independent
Borel transformation over $p^2_1$ and $p^2_2$ and only at very end
put the Borel parameters $M^2_1$ and $M^2_2$ to be equal.
 The
described procedure allows one to kill nondiagonal transitions
matrix elements of the type
\be
\langle 0 \vert j^h \vert h^* \rangle \langle h^* \vert
j^{el}_{\mu} (y) j^{el}_{\nu} (0) \vert h \rangle \langle h \vert
j^h \vert 0 \rangle \label{121}\ee and thus makes it possible to
separate the diagonal transition of interest
\be
\langle 0 \vert j^h \vert h \rangle \langle h \vert j^{el}_{\mu}
(y) j^{el}_{\nu} (0) \vert h \rangle \langle h \vert j^h \vert 0
\rangle.\label{122} \ee As was shown in Ref.\cite{A} the sum rules
for nucleon have the form
\be
\frac{2 \pi}{4 M^4}~ \frac{\bar{\lambda}^2_p}{32 \pi^4} ~ x
q^{u,d} (x) e^{-m^2/M^2} = Im T^0_{u,d} + \mbox{Power~
corrections} \label{123}\ee Here the l.h.s is the phenomenological
side of the sum rule -- the proton state contribution,
$\bar{\lambda}_p$ is defined in (100). In numerical calculations
it will be put equal $\bar{\lambda}^2_p=3.0~GeV^6$. The right hand
side is calculated in QCD. The excited states contribution -- the
continuum is identifyed with the contribution of bare loop
diagram,  starting from continuum threshold value $s_0$ and is
transferred to the l.h.s. of the sum rule. The bare loop
contribution to the sum rules is represented in Fig.14.


\begin{figure}
\hspace{43mm} \epsfig{file=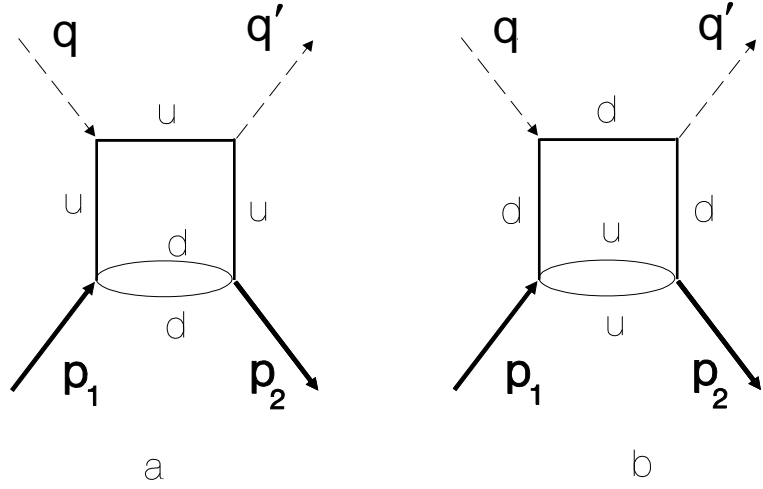, width=100mm}
\caption{Bare loop  diagrams, corresponding to unit  operator
contribution for $u$- and $d$-quarks (respectively, a) and b)).}
\end{figure}

The results after the double Borel transformation are
\be
Im T^0_{u(d)} = \varphi^{u(d)}_0 (x) \frac{M^2}{32 \pi^3}~ E_2
(s_0/M^2)\label{125} \ee where
\be
\varphi^u_0 (x) = x(1-x)^2 (1+8 x), ~~ \varphi^d_0(x) = x(1-x)^2
(1+2x), \label{126}\ee $E_2(z)$ is given by (98)
 The substitution of eq.(125)
into the sum rules (124) results in
\be
x q(x)_0^{u(d)} = \frac{2 M^6
e^{m^2/M^2}}{\bar{\lambda}^2_p}~\varphi_0^{u(d)}(x) \cdot E_2
\Biggl(\frac{s_0}{M^2}\Biggr)\label{127} \ee  Making use of
relation $\bar{\lambda}^2_p e^{-m^2/M^2} = M^6 E_2$ which follows
from the sum rule for the nucleon mass (see (95)) in the same
approximation), we get
\be\int\limits^{1}_0~d_0(x) dx = 1, ~~~ \int\limits^{1}_0~u_0(x)
dx = 2 \label{128}\ee In the bare loop approximation there also
appears the sum rule for the second moment:
\be
\int\limits^{1}_{0}~ x(q^u_0(x) + q^d_0(x)) dx = 1 \label{129}\ee

Analogously to \cite{12a} one can show that relations
(\ref{128}),(\ref{129}) hold also when taking into account power
corrections proportional to the quark condensate square in the sum
rules for the 4-point correlator   and in the sum rules for the
nucleon mass. Relations (\ref{128}) reflect the fact that proton
has two $u$-quarks and one $d$-quark. Relation (\ref{129})
expresses the momentum conservation law -- in the bare loop
approximation all momentum is carried by valence quarks.
Therefore, the sum rules (\ref{128}),(\ref{129}) demonstrate that
the zero order approximation is reasonable.

Let us calculate the perturbative corrections to bare loop and
restrict ourselves by the leading order (LO) corrections
proportional to $ln Q^2_0/\mu^2$, where $Q^2_0$ is the point,
where the quark distributions $q(x, Q^2_0)$ is calculated and
$\mu^2$ is the normalization point. In our case it is reasonable
to choose $\mu^2$ to be equal to the Borel parameter $\mu^2 =
M^2$. The results take the form:
$$ d^{LO}(x) = d_0(x) \left \{1+\frac{4}{3} ln (Q^2_0/M^2) \cdot
\frac{\alpha_s(Q^2_0)}{2 \pi} \cdot \right. $$
\be
\left. \Biggl [1/2+x+ln((1-x)^2/x) + \frac{-5-17x+16x^2+12x^3}
{6(1-x)(1+2x)} - \frac{(3-2x)x^2 ln (1/x)}{(1-x)^2(1+2x)} \Biggr ]
\right \} \label{130}\ee
$$ u^{LO}(x) = u_0(x) \cdot  \left \{
1+\frac{4}{3}~\frac{\alpha_s(Q^2_0)}{2 \pi}ln (Q^2_0/M^2) \Biggl [
1/2+x+ln(1-x)^2/x +  \right. $$
\be
\left. \frac{7-59x+46x^2+48x^3}{6(1-x)(1+8x)} - \frac{(15-8x)x^2
ln (1/x)}{(1-x)^2(1+8x)} \Biggr ] \right \} \label{131}\ee where
$u_0(x)$ and $d_0(x)$  are bare loop contributions, given by
(\ref{127}).

 The
contributions of gluon condensate to $u$ and $d$-quarks
distribution were found to be (the ratios to bare loop
contributions are presented):
\be
\frac{u(x)_{\langle G^2 \rangle}}{u_0(x)} = \frac{\langle
(\alpha_s/\pi) G^2 \rangle}{M^4} \cdot
\frac{\pi^2}{12}~\frac{(11+4x-31x^2)}{x(1-x)^2(1+8x)}~
E_0(s_0/M^2)/E_2(\frac{s_0}{M^2})\label{132} \ee
\be
\frac{d(x)_{\langle G^2 \rangle}}{d_0(x)} = - \frac{\langle
(\alpha_s/\pi) G^2 \rangle}{M^4}
\frac{\pi^2}{6}~\frac{(1-2x^2)}{x^2(1-x)^2(1+2x)}
E_0(s_0/M^2)/E_2(\frac{s_0}{M^2})\label{133} \ee The contributions
of quark condensate -- the terms, proportional to
$\alpha_s(M^2)\langle 0 \mid \bar{q} q \mid 0\rangle^2$ are few
times smaller, than the contributions of gluon condensate and are
not presented here. (They can be found in Ref.\cite{A}).


\begin{figure}[tb]
\hspace{25mm} \vspace{-10mm} \epsfig{file=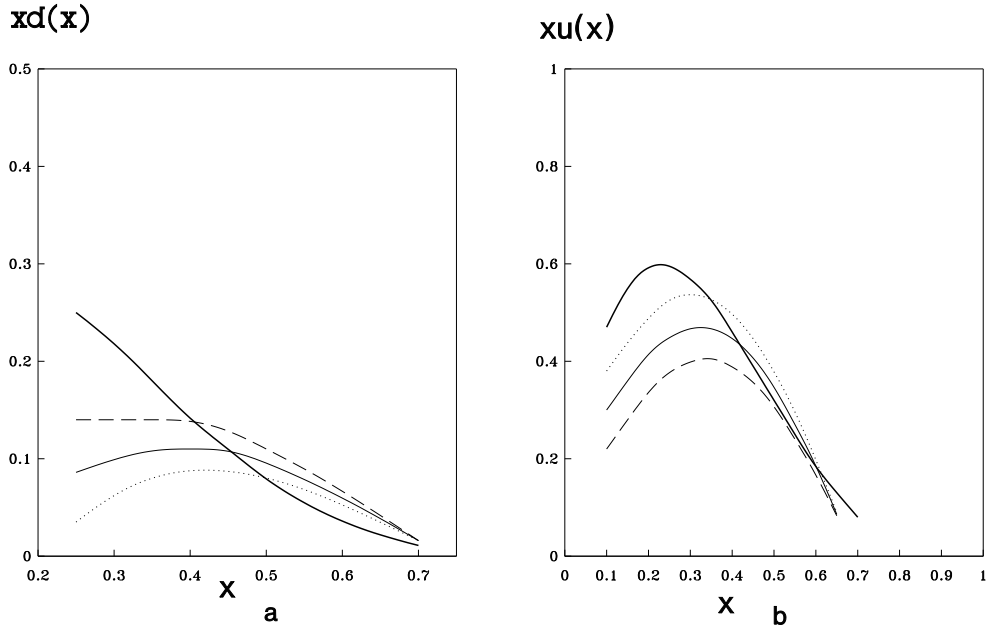, width=140mm}
\caption{$d$- and $u$-quark distribution at various values of
gluon condensate ($\langle (\alpha_s/\pi)G^2\rangle =0.012, 0.06$
and 0 GeV$^4$, respectively dotted, solid and dashed lines). Thick
solid line corresponds to the results of \cite{4a}. }
\end{figure}

The final result for valence quark distribution in proton are of
the form
$$ xu(x) = \frac{M^6 e^{m^2/M^2}}{\bar{\lambda}^2_N} 2x
(1-x)^2(1+8x)E_2(\frac{s_0}{M^2}) \left \{ \Biggl [1 +
\frac{u^{LO}(x,Q^2_0)}{u_0(x)} \Biggr ] + \right. $$
\be
\left. + \frac{1}{u_0(x)}  [ u(x)_{\langle G^2 \rangle} +
u(x)_{\alpha_s \langle \bar{q} q \rangle^2}] \right \}\label{134}
\ee $$ xd(x) = \frac{M^6 e^{m^2/M^2}}{\bar{\lambda}^2_N} 2x
(1-x)^2(1+2x)E_2(\frac{s_0}{M^2}) \left \{ \Biggl [1 +
\frac{d^{LO}(x,Q^2_0)}{d_0(x)} \Biggr ] + \right. $$
\be
\left. + \frac{1}{d_0(x)}  [ d(x)_{\langle G^2 \rangle} +
d(x)_{\alpha_s \langle \bar{q}q \rangle^2}  ] \right \}
\label{135}\ee The valence $u$ and $d$ quark distribution
calculated according to (\ref{134}),(\ref{135}) for various
$\langle (\alpha_s/\pi)G^2\rangle = 0.00, 0.006, 0.012~GeV^4$ are
shown in Fig.15.

The following values of parameters were used: $\alpha_s\langle
0\mid  \bar{q}q \mid 0 \rangle^2$ given by (\ref{85}),
$\bar{\lambda}^2_p=3.0~GeV^6$, $\lambda_{QCD}=250~GeV$,
$s_0=2.5~GeV^2$, $Q^2_0=5~GeV^2$. The contribution of $\langle
G^3\rangle$  condensate was estimated  on the basis  of instanton
model -- Eq.(\ref{29}). This contribution may influence $u$ and
$d$-quark distributions at $x \la 0.2$ and increase both of them
by 10-20\%. The  limits of applicability of QCD calculations are:
for $u$- quark -- $0.2 < x <0.65$, for $d$-quark -- $0.3 < x<
0.65$. The lower limit arises from gluon condensate contribution
-- it was required, that this contribution does  not exceed 30\%,
the upper limit was determined by increasing  of perturbative
corrections and $\alpha_s\langle \bar{q}q \rangle^2$ terms. For
comparison, Fig.15 presents the results of the fit to data basing
on the solution of DGLAP equation. The LO order fit \cite{4a} is
chosen, but not the more precise NLO fits \cite{5a}-\cite{9a},
because QCD calculations of quark distributions were performed in
LO. As is seen from Fig.15 the calculated in QCD initial (at
$Q^2_0=5~GeV^2$) valence quark distributions at $0.3 < x < 0.7$
are in a satisfactory agreement with those found from the data.
The preferred  value of gluonic condensate is $\langle 0 \mid
(\alpha_s/\pi)G^2 \mid 0 \rangle = 0.006 ~GeV^4$, the values
higher than $0.012~GeV^2$ and lower than $0$, probably, may be
excluded. These statements are in full accord with those obtained
in the previous  Chapters.

\bigskip

\section{Conclusion}

The basic parameters of QCD -- $\alpha_s(Q^2)$ and the values of
vacuum condensates, determining hadron physics at low momentum
transfers $(Q^2\sim 1-5~GeV^2)$, are reliably determined by the
theory. It was demonstrated, that the values of these parameters,
found by consideration of various processes are in a good
agreement with one another.  The values of $u,d,s$ quark masses
and their ratios are known now with a good precision -- about
10-15\% in the ratios and about 20\% in the mass absolute values.
The precision in charmed quark mass value
$\overline{m}_c(\overline{m}_c)$ in $\overline{MS}$
renormalization scheme is extremingly high -- about 1\%. The
knowledge  of $\alpha_s(Q^2)$ and condensates makes it possible to
find  the polarization operators of vector and axial currents at
$Q^2 \geq  1~GeV^2$ with high precision. In such calculation high
order perturbative terms -- $\sim\alpha^2_s$ and, in some cases,
$\sim\alpha^3_s$ must be accounted. Therefore, we have now a good
basis for theoretical description of many physical phenomena in
low energy QCD -- hadron masses, their static properties, quark
distributions in hadrons etc. Of course, at even lower momentum
transfer, $Q^2 \la 1~GeV^2$, the approach, exploited in this
review and based on perturbation theory and OPE, does not work:
the confinement mechanism and the mechanism of spontaneous
symmetry breaking  are acting in full strength. The construction
of various models is inavoidable here. But for such models the
knowledge of basic QCD parameters is also quite important -- they
may play the role of cornerstones for the models.

I summarize here the final values of $\alpha_s$ and condensates:
\be \alpha_s(m^2_{\tau}) = 0.340 \pm 0.015
~~~\overline{MS}-\mbox{scheme}\label{136}\ee
\be
\langle 0\mid \frac{\alpha_s}{\pi} G^2_{\mu\nu} \mid 0 \rangle
=0.005 \pm 0.004~GeV^2\label{137}\ee
\be
\langle 0\mid \bar{q}q\mid 0\rangle_{1~GeV} = -(1.65\pm 0.15)\cdot
10^{-2}~GeV^3,~~q=u,d\label{138}\ee
\be
\alpha_s\langle 0\mid \bar{q}q\mid 0\rangle^2 =(1.5\pm 0.2)\cdot
10^{-4}~GeV^6\label{139}\ee (In determination of (\ref{139}) the
factorization hypothesis is assumed.) The values of errors, given
in ((\ref{137})-(\ref{139}) are a bit uncertain, since the
procedure of averaging  of errors in different processes is
subjective in essential way.

\bigskip

\section{Acknowledgements}

I am thankful  to my coauthors K.N.Zyablyuk, A.G.Oganesian,
A.V.Samsonov and B.V.Geshkenbein for their cooperations in the
papers, which content was included in this review and to
N.S.Libova and M.N.Markina for their help in the preparation of
the manuscript. This work was supported in part by U.S.Civilian
Research and Development Foundation (CRDF) Cooperative Grant
Program, Project RUP2-2621-MO-04, RFBR grant 03-02-16209 and the
funds from EC to the project ``Study of Strongly Interacting
Matter'' under contract 2004 No.R113-CT-2004-506078.


\newpage
\begin{thebibliography}{99}
\bibitem{1} M.A. Shifman, A.I. Vainshtein and  V.I. Zakharov,
{\it Nucl. Phys.} B 147 (1979) 385, 448
\bibitem{2} J. Gasser and  H. Leutwyler, \Journal{\NPB}
{94} {269} {1975}
\bibitem{3} J. Gasser and H. Leutwyler, \Journal{\PREP}
{87} {77} {1982}
\bibitem{4} S. Weinberg,  in: A Festschrift for I.I. Rabi ed. by
L. Motz, Trans. New York

Acad. Sci., Ser.II, 38 (1977) p. 185
\bibitem{5} R. Dashen,  \Journal{\PREV} {183} {1245} {1969}
\bibitem{6} J. Gasser and H. Leutwyler, \Journal{\NPB} {250} {469}
{1985}
\bibitem{Perez} J. Donoghue and A. Perez, \Journal{\PRD} {55} {7075} {1997}
\bibitem{Bijnens} J. Bijnens and J. Prades, \Journal{\NPB}
{490} {293} {1997}
\bibitem{J.Gasser} J. Gasser and H. Leutwyler, \Journal{\NPB}
{250} {539} {1985}
\bibitem{Kambor} J. Kambor, C. Wiesendanger and D. Wyler, \Journal{\NPB}
{465} {215} {1996}
\bibitem{Anisovich} A.V. Anisovich and H. Leutwyler, \Journal{\PLB}
{375} {335} {1996}
\bibitem{Sopov} B.V. Martemyanov and V.S. Sopov, \Journal{\PRD} {71}
 {017501} {2005}
\bibitem{10} Particle Data Group, S. Eidelman et
al., \Journal{\PLB} {592} {1} {2004}
\bibitem{7} H. Leutwyler,  {\it J. M. Phys. Soc.} 6 (1996) 1,
hep-ph/9602255
\bibitem{8} B.L. Ioffe and M.A. Shifman, \Journal{\PLB}
{95} {99} {1980}
\bibitem{9} B.L. Ioffe and M.A. Shifman, \Journal{\PLB}
{107} {33} {1981}
\bibitem{Collab} CLEO Collaboration, N.E. Adam et al,
hep-ex/0503028
\bibitem{Nasrallah} N. Nasrallah, \Journal{\PRD}
{70} {116001} {2004}
\bibitem{11} C. Aubin et al. (MILC Collab.), hep-lat/0407028
\bibitem{12}J.A.M. Vermaseren, A.A. Larin and  T. van Ritbergen,
\Journal{\PLB} {405} {327} {1997}
\bibitem{13} S. Narison,  QCD as a Theory of  Hadrons: from Partons
to Confinement,

Cambridge Univ. Press, 2002, hep-ph/0202200
\bibitem{14} M. Jamin and A. Pich,  hep-ph/0411278
\bibitem{15} P.A. Baikov, K.G. Chetyrkin and J.H. K\"uhn,
hep-ph/0412350
\bibitem{16} G. Schierholz  et al., hep-ph/0409312
\bibitem{17} C. Adami, E.G.  Drukarev and B.L. Ioffe,
\Journal{\PRD} {48} {1441} {1993}
\bibitem{18} V.L. Eletsky and B.L. Ioffe, \Journal{\PRD}
{48} {2304} {1993}
\bibitem{Prades} J. Prades, {\it Nucl. Phys.} B (Proc. Suppl.)  64
(1998) 253
\bibitem{19}
V.A. Novikov, M.A. Shifman, A.I. Vainstein and  V.I. Zakharov,

\Journal{\NPB} {249} {445} {1985}
 \bibitem{20} M.A. Shifman,
{\it Lecture at 1997 Yukawa International Seminar},

Kyoto, 1997, Suppl.Prog.Theor.Phys., Vol. 131 (1998) p. 1
\bibitem{21} M. Gell-Mann, R.J. Oakes and B. Renner,  \Journal{\PREV}
{175} {2195} {1968}
\bibitem{22} B.L. Ioffe, {\it Physics-Uspekhi} 44 (2001) 1211
\bibitem{23} B.L. Ioffe, \Journal{\NPB} {188} {317} {1981};
Errata 192 (1982) 591
\bibitem{24} H. Leutwyler,  in:
{\it At the Frontier of Particle Physics}, Handbook of QCD,

Boris Ioffe Festschrift, ed. by M. Shifman, World Scientific, Vol.
1 (2001) p. 271
\bibitem{25}
U. Meissner, ibid, p. 417
\bibitem{26} P. Gerber and  H. Leutwyler, \Journal{\NPB} {321} {387}
{1989}
\bibitem{27} P. Chen et al., \Journal{\PRD} {64} {014503} {2001}
\bibitem{28} V.M. Belyaev and B.L. Ioffe, {\it Sov. Phys. JETP}
 56 (1982) 493
\bibitem{29} H.G. Dosch and S. Narison, \Journal{\PLB} {417} {173} {1998}
\bibitem{Giacomo} A.Di Giacomo and Yu.Simonov, hep-ph/0404044
\bibitem{30} M.A. Shifman,  {\it Pis'ma v ZhETF} 24 (1976) 376
\bibitem{31} V.A. Novikov,  M.A. Shifman, A.I. Vainstein and
V.A. Zakharov,

\Journal{\PLB} {86} {34+7} {1979}
\bibitem{32} S.N. Nikolaev and A.V. Radyushkin,  {\it Sov. J. Nucl.
Phys.} 39 (1984) 91
\bibitem{33} B.L. Ioffe and A.V. Samsonov, {\it Phys. At. Nucl.}  63
(2000) 1448
\bibitem{34} B.L. Ioffe and A.V. Smilga, \Journal{\NPB} {232} {109}
{1984}  \bibitem{35} V.M. Belyaev and I.I. Kogan,  {\it Yad. Fiz.}
40 (1984) 1035

I.I.  Balitsky, A.V. Kolesnichenko and A.V. Yung,  {\it Yad. Fiz.}
41 (1985) 282
\bibitem{36} V.M. Belyaev and I.I. Kogan, {\it Pis'ma v  ZhETF}  37 (1983)
611
\bibitem{37} B.L. Ioffe and A.G. Oganesian, \Journal{\PRD}
{57}  {R6590} {1998}
\bibitem{38} ALEPH Collaboration, R. Barate
et al., {\it Eur. Phys. J.}  C 4 (1998) 409,

 C 11 (1999) 599, The data files are taken from

http://alephwww.cern.ch/ALPUB/paper/ paper.html
\bibitem{39} OPAL Collaboration, K. Ackerstaff et al., {\it Eur. Phys. J.} C
 7 (1999) 571;

 G. Abbiendi et al., ibid,  13 (2002) 197
\bibitem{40} CLEO Collaboration,
S.J. Richichi et al., \Journal{\PRD} {60}  {112002} {1999}
\bibitem{41} B.L. Ioffe and K.N. Zyablyuk, \Journal{\NPA}
{687} {437} {2001} \bibitem{42} B.V. Geshkenbein, B.L. Ioffe and
K.N. Zyablyuk, \Journal{\PRD} {64} {093009} {2001}
\bibitem{43} M. Davier, A. H\"ocker, R. Girlanda and J. Stern,
\Journal{\PRD} {58} {96014} {1998}
\bibitem{44} M. Davier  et al., hep-ex/0312064
\bibitem{45} A. Pich,  Proc. of QCD 94 Workshop, Monpellier, 1994;
{\it Nucl. Phys.} B

 (proc.Suppl)
 39  (1995) 396
\bibitem{46} W.J. Marciano and
A. Sirlin,  \Journal{\PRL}  {61}  {1815} {1998}
\bibitem{47} E. Braaten, \Journal{\PRL}  {60} {1606} {1988};
\Journal{\PRD} {39} {1458} {1989}
\bibitem{48} S. Narison and  A. Pich,
\Journal{\PLB} {211} {183} {1988}
\bibitem{49} F. Le Diberder and A. Pich, \Journal{\PLB} {286} {147}
{1992} \bibitem{50} K.G. Chetyrkin, A.L. Kataev and F.V. Tkachov,
\Journal{\PLB} {85} {277} {1979};

M. Dine and J. Sapirshtein,  \Journal{\PRL}
 {43} {668} {1979};

 W. Celmaster and R. Gonsalves,  ibid,  44 (1980) 560
 \bibitem{51} L.R. Surgaladze and M.A. Samuel, \Journal{\PRL} {66}
{560} {1991};

S.G.  Goryshny, A.L. Kataev and  S.A. Larin,  \Journal{\PLB} {259}
{144} {1991}
\bibitem{52}
A.L. Kataev and V.V. Starshenko,  {\it Mod. Phys. Lett.}  A 10
(1995) 235
\bibitem{53} P.A. Baikov, K.G. Chetyrkin and J.P. K\"uhn, \Journal{\PRD}
{67}  {074026} {2003}
\bibitem{54} O.V. Tarasov, A.A. Vladimirov and  A.Yu. Zharkov,
\Journal{\PLB} {93} {429} {1980};

S.A. Larin and J.A.M. Vermaseren, ibid,  303 (1993) 334
\bibitem{55} T. van Ritbergen, J.A.M. Vermaseren and S.A. Larin,
\Journal{\PLB} {400} {379} {1997}
\bibitem{Czakon}M. Czakon, hep-ph/0411261
\bibitem{56} A.V. Radyushkin,  JINR E2-82-159, hep-ph/9907228
\bibitem{57} A. Pivovarov, \Journal{\NCA} {105} {813} {1992}
\bibitem{58} K.G. Chetyrkin, S.G. Gorishny and  V.P. Spiridonov,
\Journal{\PLB} {160} {149} {1985}
\bibitem{59}L.-E. Adam and
K.G. Chetyrkin,  \Journal{\PLB} {329} {129} {1994}
\bibitem{DS} M.S. Dubovikov and A.V. Smilga, \Journal{\NPB}
{185} {109} {1981}
\bibitem{60} E. Braaten and C.S. Lee, \Journal{\PRD} {42} {3888} {1990}
\bibitem{Davier} M. Davier,  8 Intern. Symposium on Heavy Flavour
Physics, Southampton,

England, 1999, hep-ex/9912094
\bibitem{61} ALEPH Collaboration, R. Barate
et al., {\it Eur. Phys. J.}  C 11 (1999) 599
\bibitem{62} OPAL Collaboration, G. Abbiendi  et al., {\it Eur. Phys.
J.} C 19 (2001) 653
\bibitem{a1} S. Bethke, hep-ex/040702
\bibitem{SShur}
T. Shafer and E.V. Shuryak, \Journal{\RMP} {70} {323} {1998}
\bibitem{EVShuryak} E.V. Shuryak, \Journal{\NPB} {198} {83} {1982}
\bibitem{S2}
"Instantons in Gauge Theories", ed. by M. Shifman, World
Scientific, 1994
\bibitem{AG}
N. Andrei and  D.J.  Gross, \Journal{\PRD} {18} {468} {1978}
\bibitem{SShur2}
T. Shafer and  E.V.  Shuryak, \Journal{\PRL} {86} {3973} {2001}
\bibitem{Luke}
Y.L. Luke,  ``Mathematical functions and their approximations'',

NY, Academic Press 1975; H. Bateman and A.  Erdelyi,  ``Higher
transcendental

 functions'', Vol.
I, Krieger, Melbourne, FL, 1953
\bibitem{68} S.J. Brodsky, G.P. Lepage and P.B.  Mackenzie, \Journal
{\PRD} {28} {228} {1983}
\bibitem{69} S.J. Brodsky, V.S. Fadin, V.T. Kim, L.N. Lipatov and
G.B. Pivovarov,

 {\it JETP Lett.} 70 (1999) 155
\bibitem{70} S.J. Brodsky, S. Menke, C. Merino and J. Rathsman,
\Journal{\PRD} {67} {055008} {2003}
\bibitem{71} D.V. Shirkov,  {\it Eur. Phys. J.} C 22 (2001) 331
\bibitem{72} K.A. Milton, I.L. Solovtsov and O.P. Solovtsova,
\Journal{\PRD} {64} {016005} {2001}
\bibitem{Dubovikov} M.S.  Dubovikov and A.V. Smilga,  {\it Yad. Fiz.}
37 (1983) 984

A. Grozin and Y. Pinelis, \Journal{\PLB} {166} {429} {1986}
\bibitem{76} K.N. Zyablyuk,  hep-ph/0404230.
\bibitem{77} K.G. Chetyrkin, S.G. Gorishny and V.P. Spiridonov,
\Journal{\PLB}  {160} {149} {1985}
\bibitem{78} L.-E. Adam and
K.G. Chetyrkin, \Journal{\PLB} {329} {129} {1994}
\bibitem{b} J. Bijnens, E. Gamiz and J. Prades, {\it JHEP} 10 (2001)
009
\bibitem{c} V. Cirigliano, E. Golowich and K.Maltman,
\Journal{\PRD} {68} {054013} {2003}
\bibitem{d} S. Friot, D. Greynat and E. de Rafael, {\it JHEP} 10 (2004)
043
\bibitem{e} J. Rojo and J. I. Lattore, {\it JHEP} 0401 (2004) 055
\bibitem{f} C.A. Dominguez and A. Schilcher, \Journal{\PLB} {581}
{193} {2004}
\bibitem{g} S. Chiulli, G. Sebu, K. Schilcher and G. Speiberger,
\Journal{\PLB} {595} {359} {2004}
\bibitem{h} S. Narison, hep-ph/0412152
\bibitem{79} ALEPH Coollaboration, R. Barate,
et al., \Journal{\ZPC} {76} {15} {1997}
\bibitem{a2}S.I. Eidelman, L.M. Kurdadze and A.I. Vainstein,
\Journal{\PLB} {82}  {278} {1979}

\bibitem{Hocker} A. H\"ocker, hep-ph/0410081
\bibitem{80} B.L. Ioffe, \Journal{\ZPC} {18} {67} {1983}
\bibitem{81} M. Jamin,  \Journal{\ZPC} {37} {635} {1988};

M. Jamin, Dissertation  thesis, Heidelberg preprint HD-THEP-88-19
(1988);

A.A.  Ovchinnikov, A.A. Pivovarov and L.R. Surguladze, {\it Sov.
J. Nucl. Phys.} 48 (1988) 358;

A.G. Oganesian, hep-ph/0308289
\bibitem{Dorokhov} A.E. Dorokhov and N.I. Kochelev,
 \Journal{\ZPC} {46}
{281} {1990}
\bibitem{Forkel} H. Forkel and M.K. Banerjee, \Journal{\PRL} {71}
{484} {1993}
\bibitem{82} S.N. Nikolaev and A.V. Radyushkin,  {\it JETP Lett.}
37 (1982) 526
\bibitem{83} L.J. Reinders, H.R.  Rubinstein and S. Yazaki,
\Journal{\PLB} {138}  {340} {1984}
\bibitem{84}
S. Narison,  QCD Spectral Sum Rules, (World Scientific, 1989);

\Journal{\PLB} {387} {162} {1996};

V.A. Novikov, M.A. Shifman, A.I. Vainstein, M.B.  Voloshin and
V.A. Zakharov,

\Journal{\NPB} {237} {525} {1984};

K.J. Miller and M.G.  Olsson,  \Journal{\PRD} {25} {1247} {1982}
\bibitem{85}
P. Colangelo and A. Khodjamirian,  At frontier of Particle
Physics, Handbook

of QCD, Boris Ioffe Festschrift, v.3, p.1495, World Scientific,
2001.
\bibitem{86}
R.A. Bertlmann,  \Journal{\NPB} {204} {387} {1982};

V.N. Baier and Yu.F.  Pinelis,  \Journal{\PLB} {116} {179} {1982},

\Journal{\NPB} {229} {29}  {1983};

G. Launer, S. Narison and R. Tarrach, \Journal{\ZPC} {26} {433}
{1984};

R.A. Bertlmann, C.A.  Dominguez, M. Loewe, M. Perrottet and E. de
Rafael,

\Journal{\ZPC} {39} {231}  {1988};

P.A. Baikov, V.A.  Ilyin and V.A.  Smirnov,  {\it Phys. Atom.
Nucl.} 56 (1993) 1527;

D.J. Broadhurst, P.A. Baikov, V.A. Ilyin, J. Fleischer, O.V.
Tarasov and

V.A.  Smirnov, \Journal{\PLB} {329} {103} {1994};

B.V. Geshkenbein,  {\it Phys. Atom. Nucl.}  59 (1996) 289
\bibitem{87} M. Jamin and A. Pich,  \Journal{\NPB} {507} {334} {1997};

M. Eidem\"uller and M.  Jamin, \Journal{\PLB} {498} {203} {2001};

J.N. K\"uhn and M.  Steinhauser, \Journal{\NPB} {619} {588}
{2001}
\bibitem{88}
B.L. Ioffe and K.N. Zyablyuk,  {\it Eur. Phys. J.} C 27  (2003)
229
\bibitem{89} V.B. Berestetsky and I.Ya. Pomeranchuk, {\it JETP} 29
(1955) 864
\bibitem{90} J. Schwinger, Particles, Sources, Fields, Addison-Wesley
Publ., 1973, V.2 \bibitem{91} A.H. Hoang, J.H. K\"uhn and T.
Teubner, \Journal{\NPB} {452} {173} {1995};

K.G. Chetyrkin, J.H. K\"uhn and M. Steinhauser, \Journal{\NPB}
{482} {231} {1996};


K.G. Chetyrkin et al.,  \Journal{\NPB}  {503} {339}  {1997};

K.G. Chetyrkin et al., {\it Eur. Phys. J.} C 2 (1998) 137

\bibitem{92} D.J. Broadhurst et al., \Journal{\PLB}  {329}
{103} {1994} \bibitem{93} S.N. Nikolaev and A.V. Radyushkin, {\it
Sov. J. Nucl. Phys.} 39 (1984) 91
\bibitem{94} K.N. Zyablyuk,  {\it JHEP} 0301:081 (2003)
\bibitem{95} A.V Samsonov,  hep-ph/0407199
\bibitem{96} V.A. Khoze and M.A. Shifman,
{\it Sov. Phys. Usp.}  26 (1983) 387
\bibitem{97} M.B. Voloshin,  {\it Int. J. Mod. Phys.} A 10
(1995) 2865
\bibitem{98}
M. Jamin, A.  Pich, \Journal{\NPB} {507} {334} {1997}
\bibitem{99} K.G. Chetyrkin, A.H.  Hoang, J.H. K\"uhn,
M. Steinhauser and T.  Teubner,

{\it Eur. Phys. J.} C 2 (1998) 137
\bibitem{100} J.H. K\"uhn, A.A. Penin and A.A.  Pivovarov,  \Journal{\NPB}
{534} {356} {1998}
\bibitem{101} L. Landau and  E. Lifshitz,
Quantum Mechanics: Nonrelativistic Theory,

Pergamon Press, 1977
\bibitem{102} E. Eichten et al., \Journal{\PRD} {21} {203} {1980}
\bibitem{103} B.V. Geshkenbein,  \Journal{\PRD} {70} {074027}
{2004}
\bibitem{1a} V.N. Gribov and L.N. Lipatov, {\it Sov. J. Nucl.
 Phys.} 15  (1972) 438
\bibitem{2a} Yu.L. Dokshitzer, {\it Sov. Phys. JETP} 46  (1977) 641
\bibitem{3a} G. Altarelli and G. Parisi, \Journal{\NPB} {126} {298}
{1977}
\bibitem{4a} M. Gl\"uck, E. Reya and A. Vogt, \Journal{\ZPC} {53} {127}
{1992}
\bibitem{5a} H. L. Lai et al. (CTEQ Collab.), {\it Eur. Phys. J.}
C 12 (2000) 375
\bibitem{6a} A.D. Martin, R.G. Roberts,
 W.J. Stirling and R.S. Thorne, {\it Eur. Phys.

J.} C 4 (1998) 463
\bibitem{7a} M. Gl\"uck, E. Reya and A. Vogt,
{\it Eur. Phys. J.} C 5 (1998) 461
\bibitem{8a} A.M. Cooper-Sarkar, R.C.E. Davenish and A. De Roeck,
{\it Int. J. Mod. Phys.}

A 13 (1998) 3385
\bibitem{9a} Wu-Ki Tung, At the Frontier of Particle Physics, in: Handbook
of QCD,

Boris Ioffe Festschrift, ed. by M. Shifman, World Sci. 2001, v.2,
p.887
\bibitem{11a} B.L. Ioffe, {\it Pis'ma v ZhETF}
 42 (1985) 266, JETP Lett.  42 (1985) 327
\bibitem{12a} V.M. Belyaev and B.L. Ioffe, \Journal{\NPB} {310} {548} {1988}
\bibitem{13a} A.S. Gorsky, B.L. Ioffe, A.Yu.
 Khodjamirian and A.G. Oganesian,

\Journal{\ZPC} {44} {523} {1989}

B.L. Ioffe and A.G.Oganesian, \Journal{\ZPC} {69} {119} {1995}
\bibitem{14a} B.L. Ioffe and A.Yu. Khodjamirian, \Journal{\PRD}
{51} {3373} {1995}
\bibitem{15a} B.L. Ioffe and A. Oganesian, {\it Eur. Phys. J.}
C 13 (2000) 485
\bibitem{10a} B.L. Ioffe and A.G. Oganesian,
\Journal{\PRD} {63} {096006} {2001}
\bibitem{16a} V.N. Gribov, B.L. Ioffe and I.Ya. Pomeranchuk,
{\it Sov. J. Nucl. Phys.} 2 (1966) 549
\bibitem{17a} B.L. Ioffe, \Journal{\PLB} {30}  {123} {1969}
\bibitem{A} B.L. Ioffe, A.G. Oganesian,
\Journal{\NPA} {714} {145} {2003}




\end{thebibliography}
\end{document}